\documentclass[preprint1]{aastex}
\usepackage{graphicx}
\usepackage{epsfig}
\usepackage{amssymb,amsmath}

\slugcomment{To appear in ApJ}
\shorttitle{The Nature of  Transition Disks}
\shortauthors{Cieza et al.}

\begin{document}

\title{The Nature of  Transition Circumstellar Disks I. \\ The Ophiuchus Molecular Cloud{\LARGE{$^{\star}$}}}

\author{Lucas A. Cieza\altaffilmark{1,}\altaffilmark{2},
Matthias R. Schreiber\altaffilmark{3},
Gisela A. Romero\altaffilmark{3,4},
Marcelo D. Mora\altaffilmark{3},
Bruno Merin\altaffilmark{5},
Jonathan J. Swift\altaffilmark{1},
Mariana Orellana\altaffilmark{3,4},
Jonathan P.  Williams\altaffilmark{1},
Paul M. Harvey\altaffilmark{6},
Neal J. Evans II\altaffilmark{6}
}

\altaffiltext{}{$\star$ Based in part on observations made with ESO telescopes at Paranal Observatory, under ESO program 083.C-0459(A).}
\altaffiltext{1}{Institute for Astronomy, University of Hawaii at Manoa,  Honolulu, HI 96822.} 
\altaffiltext{2}{\emph{Spitzer} Fellow, lcieza@ifa.hawaii.edu}
\altaffiltext{3}{Departamento de Fisica y Astronomia, Universidad de Valpara\'{\i}so, Valpara\'{\i}so, Chile}
\altaffiltext{4}{Facultad  de Ciencias Astron\'omicas y Geof\'{\i}sicas, UNLP, La Plata, Argentina}
\altaffiltext{5}{European Space Agency (ESAC), Villanueva de la Canada, Madrid, Spain}
\altaffiltext{6}{Department of Astronomy, University of Texas at Austin,  Austin, TX 78712}
\begin{abstract}

We have obtained millimeter wavelength photometry, high-resolution
optical spectroscopy and adaptive optics near-infrared imaging for a sample of  26
 \emph{Spitzer}-selected transition circumstellar disks. All of our targets are located in 
 the Ophiuchus molecular cloud  (d $\sim$125 pc) and have Spectral Energy Distributions 
(SEDs) suggesting  the presence of inner opacity holes.
We use these ground-based data to estimate  the disk mass,  multiplicity, and 
accretion rate for each object in our sample in order to  investigate the mechanisms 
potentially responsible for  their inner holes.
We find that transition disks are a heterogeneous group of objects, with
disk masses ranging from  $<$ 0.6 to 40 M$_{JUP}$ and accretion rates 
ranging from $<$10$^{-11}$ to 10$^{-7 }$ M$_{\odot}$yr$^{-1}$,   
but most tend to have much lower masses and accretion rates than ``full disks'' 
(i.e., disks without opacity holes).  
Eight of our targets have stellar companions: 6 of them are binaries and the other 2 
are triple  systems. In four cases,  the  stellar companions are close enough to suspect
they are responsible for the inferred inner holes. 
We find that  9 of our 26 targets have low disk mass ($<$ 2.5 M$_{JUP}$) 
and negligible accretion ($<$ 10$^{-11}$ M$_{\odot}$yr$^{-1}$), and are 
thus consistent with photoevaporating (or photoevaporated) disks. 
Four of these 9 non-accreting objects have fractional disk 
luminosities $<$ 10$^{-3}$ and could already be in a debris disk stage.
Seventeen of our transition disks are accreting.  
Thirteen of these  accreting objects are consistent with  grain growth.
The remaining 4  accreting objects have SEDs suggesting the 
presence of sharp inner holes, and thus are  excellent candidates for 
harboring giant planets.

\end{abstract}
\keywords{circumstellar matter ---  binaries: general --- planetary systems: protoplanetary disks ---
stars: pre-main sequence}

\section{Introduction}\label{intro}

Multi-wavelength observations of nearby star-forming regions have shown that the vast majority of 
pre-main-sequence (PMS) stars are either accreting Classical T Tauri Stars (CTTSs)  with excess 
emission extending all the way from the   near-IR to the millimeter or, more evolved, non-accreting, 
Weak-line T Tauri Stars  (WTTSs) with bare stellar photospheres.  The fact that very few objects 
lacking near-IR excess show mid-IR or (sub)millimeter excess emission implies that, once the 
inner disk dissipates, the entire primordial disk  disappears very rapidly (Wolk $\&$ Walter 1996; Andrews $\&$ 
Williams 2005,2007; Cieza et al.  2007). 
The few objects that are caught in the short transition between typical CTTSs 
and disk-less WTTSs usually have optically thin or non-existent inner disks 
and optically thick outer disks (i.e., they have reduced opacity in the inner 
regions of the disk). 

The reduced opacity in the inner disk is the defining characteristic of the so called  transition disks.
The precise definitions of what constitutes a transition object found in the disk evolution literature are,  however,  far 
from homogeneous (see Evans et al. 2009 for a detailed description of the transition disk nomenclature). 
Transition disks have been defined as objects with no detectable near-IR excess, 
steep slopes in the mid-IR, and large far-IR excesses (e.g., Muzerolle et al. 2006, 
Sicilia-Aguilar et al. 2006a)
This definition has been 
relaxed by some authors (e.g.,  Brown et al. 2007, Cieza et al. 2008) to include objects  with small, 
but still detectable, near-IR excesses. 
Transition disks have also been more broadly defined in terms of a significant decrement 
relative to the Taurus median Spectral Energy Distribution (SED) at any or all wavelengths 
(e.g., Najita et al. 2007). This broad definition is the closest one to the criteria 
we adopt to select our sample (see \S~\ref{selection}).  Even though,  according to our definition,
 many transition disks have inner disks with non-zero opacity, we refer to their regions 
of low opacity as the ``inner opacity hole."
Recent submillimeter studies provide dramatic support for the presence of inner 
opacity holes inferred from the SED modeling of transition disks. High resolution 
submillimeter continuum images of objects such as LkH$\alpha$ 330 (Brown et al. 2008) 
and GM Tau (Hughes et al. 2009) show sharp inner holes tens of AU in radius and 
lend confidence to the standard interpretation of transition disk SEDs. 

One of the most intriguing results from transition  disk studies has been the
great diversity of SED morphologies revealed by the \emph{Spitzer} Space Telescope.
In an attempt to describe distinct classes of transition disks, several names have recently emerged in 
the literature (Evans et al. 2009). These new names include: anemic disks, flat disks, or homologously 
depleted disks to describe objects whose observed SEDs are significantly below the median of the CTTS 
population at all IR wavelengths (Lada et al. 2006, Currie et al. 2009). These new names also include 
cold disks, objects with little or no near-IR excesses whose SEDs raise very steeply  in the mid-IR 
(Brown et al. 2007), and pre-transitional disks, disks with evidence for an optically-thin gap separating 
optically-thick inner and outer disk components (Espaillat et al. 2008).


Studying the diverse population of transition disks is key  for understanding circumstellar disk evolution as much of 
the diversity of their SED  morphologies is likely to arise from different physical  processes dominating the 
evolution of different disks.  Disk evolution processes include:  viscous accretion (Hartmann et al. 1998), photoevaporation 
(Alexander et al., 2006), the magnetorotational instability (Chiang $\&$ Murray-Clay, 2007),  grain growth and 
dust settling (Dominik $\&$ Dullemond,  2008),  planet formation (Lissauer, 2003;  Boss et al. 2000), 
and dynamical interactions between the disk and stellar or substellar companions (Artymowicz  $\&$ Lubow, 1994).  
Understanding the relative importance of these physical processes in disk evolution and their
connection to the different classes of transition disks is currently one of the main 
challenges of the disk evolution field.  
Also,  even though transition disks are rare, it is likely that all circumstellar disks go through a short transition 
disk stage (as defined by their SEDs) at some point of their evolution. 
This is so because observations show that an IR excess at a given wavelength is \emph{always}  accompanied 
by an excess at longer wavelengths (the converse is not true as attested by the very existence of transition disks).  
This implies that, unless some disks manage to lose the near-, mid- and far-IR excess at  \emph{exactly} 
the same time, the near-IR excess   \emph{always} dissipates before the mid-IR and far-IR excess do.  
Since no known process is expected to remove the circumtellar dust at all radii simultaneously,  it is 
reasonable to conclude that transition disks represent a common (if not unavoidable)  phase in the evolution 
of a circumstellar disk.

At least four different mechanisms  have been proposed  to explain the opacity holes of transition 
disks: giant planet formation, grain growth, photoevaporation, and tidal truncation in close binaries.
However, since all these processes can in principle result in similar  IR SEDs,
additional observational constraints are necessary to distinguish among them. 
As discussed by Najita et al. (2007), Cieza (2008), and  Alexander (2008), 
disk masses, accretion rates, and multiplicity information are particularly 
useful to distinguish between  the different  mechanisms that are likely to produce the inner holes in transition disks.
A vivid example of the need for these kind of  data is the famous disk around CoKu Tau/4.
Since its sharp inner hole was discovered by \emph{Spitzer}  (Forrest et al. 2004), its origin has been a matter 
of great debate.  The hole  was  initially modeled to be  carved by a giant planet (Quillen et al. 2004), but Najita et 
al. (2007) argued that the low  mass and low accretion rate of the CoKu Tau/4 disk are more consistent with  photoevaporation 
than with a planet formation scenario.  More recently, Coku Tau/4 has has been shown to be a close binary star  system 
with a 8 AU projected separation, rendering its disk a circumbinary one (Ireland $\&$ Kraus 2008).

Since the number of well characterized transition disks is still in the tens, 
most studies so far have focused on the modeling of  individual objects such as
TW Hydra, (Calvet et al. 2004),  GM Aur  and DM Tau (Calvet et al. 2005),  
LkH$\alpha$ 330 (Brown et al. 2008), and LkCa 15 (Espaillat et al. 2008). 
To date, few studies have discussed the properties of transition disks as a group (e.g., 
Najita et al. 2007; Cieza et al.  2008).  These papers studied relatively small samples, suffer 
from different selection biases, and arrived, not too surprisingly, at very different conclusions.   
Najita et al. (2007) studied a sample of 12  transition objects in Taurus and found that they 
have stellar accretion rates $\sim$10 times lower and a median disk mass
($\sim$25 M$_{JUP}$) that is $\sim$4 times larger than the rest of the disks in Taurus.
They argue that most of the transition disks in their sample are consistent with the planet
formation scenario. The disk masses found by Najita et al. are in stark contrast to
the results from the  SMA study  of 26 transition disks from  Cieza et al.  (2008).  They 
observed mostly WTTSs disks and found  that \emph{all}  of them have very low masses $<$1-3 M$_{JUP}$
suggesting that  their inner holes were more likely due to photoevaporation, instead of 
the formation of jovian planets. This discrepancy can probably be traced back
to the different sample selection criteria as Najita et al. studied mostly CTTSs, while  Cieza et al. studied 
mostly WTTSs (see \S~\ref{m_a_sec}).   
A  much larger and unbiased sample of transition disks is needed to quantify the importance of 
multiplicity, photoevaporation, grain growth, and planet formation
on the evolution of circumstellar disks.

This paper is the first  part of a series from  an ongoing project aiming to characterize over 
100  \emph{Spitzer}-selected transition disks located in nearby star-forming regions. 
Here we present millimeter wavelength photometry (from the SMA and the CSO),  
high-resolution optical spectroscopy (from the Clay, CFHT, and Du Pont telescopes), 
and Adaptive Optics  near-infrared imaging  (from the VLT)  for a sample of  
26 \emph{Spitzer}-selected transition circumstellar disks  located in the Ophiuchus molecular cloud.
We use these new ground-based data to estimate  the disk mass,  accretion rate,
and multiplicity for each object in our sample
in order to  investigate the mechanisms  potentially responsible for their inner opacity holes.  
The structure of this paper is as follows.  Our  sample selection criteria are presented in \S~2, while our 
observations and data reduction techniques are described in \S~3. We present our results on disk masses, 
accretion rates, and multiplicity in \S~4. In \S~5,  we discuss the properties of our  transition disk sample and 
compare them  to those of non-transition objects.   We also discuss the likely origins of the inner holes
of  individual targets  and the implications of our results for disk evolution. Finally, a summary of our 
results and conclusions is presented in \S~6.

\section{Sample Selection}\label{selection}

\begin{figure}
\includegraphics[width=4.5in]{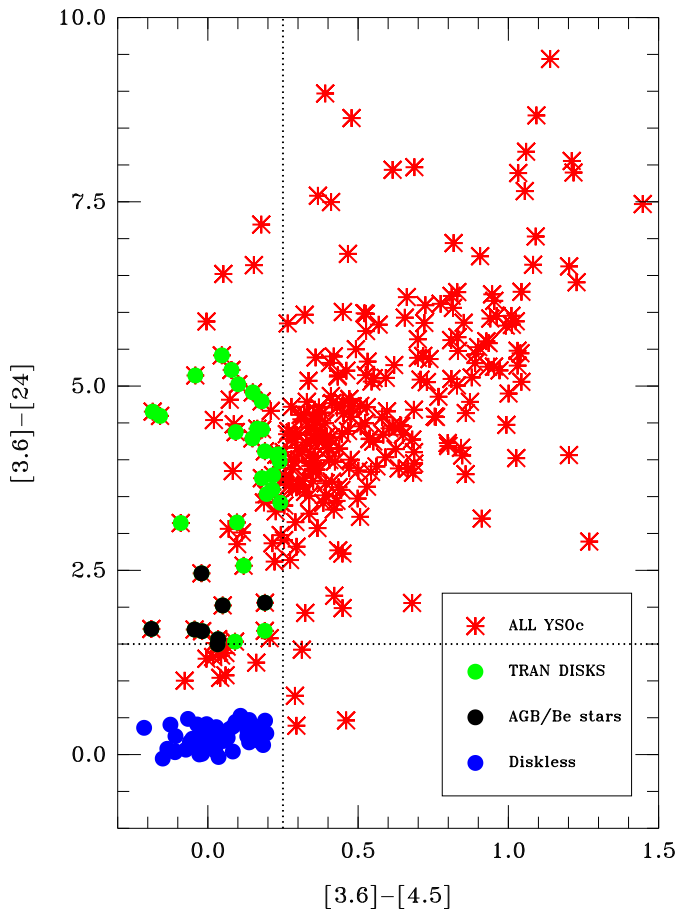}
\caption{
Color-color diagram  illustrating our key target selection criteria.
Objects with [3.6]-[4.5] $<$ 0.25 and [3.6] -[24] $<$ 0.5 are consistent
with bare stellar photospheres. Blue dots are disk-less WTTSs from
Cieza et al. (2007) used to define this region of the diagram.
Red stars are all the 297 Young Stellar Objects Candidates (YSOc) in the Cores to Disks catalog 
of  Ophiuchus.  Most PMS stars are either disk-less or have excesses at both 4.5 and 24 $\mu$m.
Our 26 transition disks, shown as green dots,   have significant ($>$ 5-$\sigma$) excess
at 24 $\mu$m and little or no excess at 4.5 $\mu$m,  as expected for disks with inner holes. 
The 8 black dots are likely Be and AGB stars contaminating our original sample.}
\label{sample_sel}
\end{figure}

We drew our sample from the 297  Young Stellar Object Candidates (YSOc) in 
the Ophiuchus catalog\footnote{available at the Infra-Red Science Archive \tt
http://irsa.ipac.caltech.edu/data/SPITZER/C2D/} of the \emph{Cores to Disks} (Evans et al. 2003) 
\emph{Spitzer} Legacy Project. For a description of the \emph{Cores to Disks}  data products, 
see Evans et al. (2007) \footnote{available at \tt http://irsa.ipac.caltech.edu /data /SPITZER/C2D/doc}.
In particular,  we selected all the targets  meeting the following criteria:  
 
\noindent{\emph{a.}}   Have \emph{Spitzer} colors [3.6]-[4.5]  $<$  0.25.  These YSOc are objects with small or 
no near-IR excess (see Figure~\ref{sample_sel}). The lack of a  [3.6] - [4.5]  color excess in our targets is 
inconsistent with an optically thick  disk extending inward to the dust sublimation radius,  and therefore indicates 
the presence of an inner opacity hole. The presence of this inner opacity hole is the defining feature we 
intend to capture in our sample.  This feature is present in $\sim$21$\%$ of the YSOc in Ophiuchus as our 
first criterion selects 63 of them (out of 297). 

\noindent{\emph{b.}}  Have \emph{Spitzer} colors  [3.6]-[24] $>$ 1.5.  We apply this criterion to ensure all our targets 
have very significant excesses. It removes the 10 YSOc  with smallest 24 $\micron$ excess 
(i.e.,  1.5 $>$  [3.6]-[24]  $>$ 1.0) and leaves 53 targets. 

\noindent{\emph{c.}}   Are detected with signal to noise ratio $>$ 7 in all  2MASS and  IRAC wavelengths as 
well as at 24 $\mu$m, to ensure we only include targets with very reliable photometry. This criterion
removes 9 objects and leaves 44 targets.

\noindent{\emph{d}}.   Have K$_S$ $<$ 11 mag ,  driven by the sensitivity of our near-IR  adaptive optics observations 
and to ensure a negligible extragalactic contamination (Padgett et al. 2008). This criterion removes 6 objects
and leaves 38 targets. 

\noindent{\emph{e.}}   Are brighter than R  = 18 mag according to the USNO-B1 catalog  (Monet et al. 2003),
driven by the sensitivity of our optical spectroscopy observations.  This  criterion removes 4 objects 
and leaves a final target list of 34 YSOc.

The first  two selection criteria ( [3.6]-[4.5]  $<$  0.25 and [3.6]-[24] $>$ 1.5 ) effectively become  our
working definition for a transition disk. These criteria are fairly inclusive and encompass most of 
the transition disk definitions discussed in \S~\ref{intro} as they select disks  with a significant  flux decrement  relative to 
 ``full disks"  in the near-IR or at all  wavelengths.
Many of our targets have IR excesses that  only become evident at 24 $\mu$m (see \S ~\ref{sed_mor}).   
To check the reality of these excesses, we have visually inspected the  cutouts of the 24 $\mu$m Ophiuchus  mosaic 
created by the \emph{Cores to Disks} project\footnote{ available at  \tt http://irsa.ipac.caltech.edu/data/SPITZER/C2D/index$\_$cutouts.html}
to confirm they show bona fide detections of our targets.  We have also verified that all the 24 $\mu$m   images of 
our targets have been assigned an  ``Image Type" = 1 in the  \emph{Cores to Disks} catalogs, corresponding to objects
that are well fitted by a point source profile.  As a final check, we have  verified that all our targets  have 
``well behaved" SEDs that are consistent with  reddened  stellar photospheres shortward  of 4.5 $\mu$m 
and IR excesses from a disk at longer wavelengths.
We have observed all 34 YSOc in our target list. However, as discussed in \S~\ref{pms_id}, 
this list includes one likely classical Be star and 7 likely Asymptotic Giant Branch stars. The remaining 26 targets 
are bona fide PMS stars with circumstellar disks and constitute our science sample. 
The \emph{Spitzer} and alternative names names, 2MASS and \emph{Spitzer} fluxes, and  the USNO-B1 R-band
magnitudes for all our 34 targets are listed in Table 1. 

\section{Observations}

\subsection{Millimeter Wavelength Photometry }

Two of our 34 targets,  \#14 and 17,  have already been detected at  millimeter wavelength 
(Andrews \& Williams, 2007 ), while stringent upper limits  exist for 3 others,  \#12, 13, and 27  
(Cieza et al. 2008).  
We have observed 24 of the 29 remaining objects with the Submillimeter Array (SMA; Ho et al. 2004),
and 5 of them with Bolocam at the Caltech Submillimeter Observatory (CSO).
In \S~\ref{disk_mass_sec}, we use the millimeter wavelength photometry to constrain the masses 
of our transition disks.

\subsubsection{Submillimeter Array Observations}

Millimeter interferometric observations of  24 of  our targets were conducted in service mode 
with the SMA, on Mauna Kea, Hawaii,  during the Spring and Summer of  2009 
(April 6$^{th}$ through July 16$^{th}$)  in the compact-north configuration and with the  
230 GHz/1300 $\mu$m receiver.
Both the upper and lower sideband data were used, resulting in a total bandwidth of 4\,GHz.

Typical zenith opacities during our observations were $\tau_{225\,{\rm GHz}}$  $\sim$0.15--0.25.
For each target, the observations cycled between the target and
two gain calibrators, 1625-254 and 1626-298, with 20-30 minutes on target and
 7.5 minutes on each calibrator. 
The raw visibility data were calibrated with the MIR reduction
package\footnote{available at  \tt  http://cfa-www.harvard.edu/$\sim$cqi/mircook.html}.
The passband was flattened using $\sim$1 hour scans of  3c454.3
and the solutions for the  antenna-based complex gains  were obtained using 
the primary calibrator 1625-254.  These gains, applied to our secondary calibrator, 1626-298, 
served as a consistency check for the solutions.

The absolute flux scale was determined through   observations of either  
Callisto or Ceres and is estimated to be accurate to 15$\%$.
The flux densities of detected sources were measured by fitting a point source model
to the visibility data, while upper limits were derived from the rms of the
visibility amplitudes. The rms noise of our SMA observations range from 1.5 to 5 mJy per beam. 
We detected, at the 3--$\sigma$ level or better,  5 of our 24 SMA targets: \#3, 15, 18, 21, and 32.
The 1.3 mm fluxes (and 3-sigma upper limits) for our entire SMA  sample are 
listed in Table 2.

\subsubsection{CSO-Bolocam Observations}

Millimeter wavelength observations of 5 of our targets were made with 
Bolocam\footnote{ \tt http://www.cso.caltech.edu/bolocam/} at the CSO on Mauna Kea, 
Hawaii, during  June 25$^{th}$-30$^{th}$, 2009. The observations were performed in the 1.1 mm mode, which 
has a bandwidth of 45 GHz centered at 268 GHz.  
Our sources were scanned using a Lissajous pattern providing 10 min of integration time per scan.
Between 16 and 40  scans per source were obtained. The weather was 
clear for the run, with $\tau_{225\,{\rm GHz}}$ ranging from  0.05 to  0.1. 
Several quasars close to the science fields were used as poniting calibrators,
while Uranus and Neptune were used as flux calibrators. The data were reduced and 
the maps from individual scans were coadded using the Bolocam analysis pipeline 
\footnote{ \tt http://www.cso.caltech.edu/bolocam/AnalysisSoftware.html}, 
which consists of a series of modular IDL routines.  None of the 5 targets were detected by 
Bolocam in the coadded maps, which have rms noises ranging from 3 to  6 mJy  per beam.
The 3-$\sigma$ upper limits for the 1.1 mm fluxes of our Bolocam targets are listed in Table 
2.

\subsection{Optical Spectroscopy}

We obtained Echelle spectroscopy  (resolution $>$ 20,000) for our entire sample using 3 different telescopes,  Clay, CFHT, and Du Pont. All the spectra include the H$\alpha$ line,  which we use to 
derive accretion rates  (see \S ~\ref{Acc}).

\subsubsection{Clay--Mike Observations}

We observed 14  of our targets with the Magellan Inamori Kyocera Echelle (MIKE)
spectrograph on the 6.5-m Clay telescope at Las Campanas Observatory, Chile.
The observations were performed in visitor mode on April 27$^{th}$ and 
28$^{th}$, 2009. 
We used the red arm of the spectrograph and a 1$''$ slit to  obtain  
complete optical spectra between 4900 and 9500  \AA \   
at a  resolution of 22,000. This resolution corresponds to $\sim$0.3 \AA \ 
at the location of the  H$\alpha$  line and to a  velocity dispersion of $\sim$14 km/s.  
Since the CCD of MIKE's red arm has a pixel scale of 0.05 \AA/pixel, we binned
the detector by a factor of 3 in the dispersion direction and  a factor of 2
in the spatial direction in order to reduce the readout time and noise. 
The R-band magnitudes of our MIKE  targets range from  15.5  to  18.
For each object,  we obtained a set of 3 or 4 spectra, with exposure 
times ranging from 3 to 10 minutes each, depending of the brightness of 
the targets.  The data were reduced using the standard IRAF packages  
IMRED:CDDRED and  ECHELLE:DOECSLIT. 

\subsubsection{CFHT--Espadons Observations}

Twelve of our targets were observed with the ESPaDonS Echelle spectrograph on
the 3.5-m Canada-France-Hawaii Telescope (CFHT) at Mauna Kea
Observatory. The observations were performed in service mode 
during the last ESPaDonS observing run of the 2009A semester 
(June 30$^{th}$-July 13$^{th}$).
The spectra were obtained in the standard ``star+sky" mode,
which delivers the complete optical spectra between 3500 \AA \
and 10500 \AA \  at a resolution of 68,000, or 4.4 km/s. 
The R-band magnitudes of our ESPaDonS  targets range between 7 and 15.
For each object, we obtained  a set of 3 spectra with exposures times
ranging from 2.5 to 10 minutes each, depending on the brightness
of the target. The data were reduced through the standard CFHT pipeline 
Upena, which is based on the  reduction package 
Libre-ESpRIT\footnote{ \tt http://www.cfht.hawaii.edu/Instruments/Spectroscopy/Espadons/Espadons\_esprit.html}.

\subsubsection{Du Pont--Echelle  Observations}

We observed 8  of our targets with the Echelle Spectrograph on the 2.5-m 
Irenee du Pont telescope at las Campans Observatory.  The observations were performed 
in visitor mode between May 14$^{th}$ and May 16$^{th}$, 2009.
We used a 1$''$ slit to obtain spectra between 4000 and 9000 \AA \  with a resolution of 45,000, 
corresponding to  0.14 \AA~in the red. However,  since the Spectrograph's CCD has a pixel scale of 
$\sim$0.1 \AA/pixel, the true two-pixel resolution corresponds to $\sim$32,000, or $\sim$9.4 km/s 
in the red.  The R-band magnitudes of our Du Pont targets range between 15.0 and 16.5.
For each object, we obtained  a set of 3 to 4  spectra with exposures times
ranging from 10  to 15 minutes each, depending on the brightness of the target. 
The data reduction was performed using the standard IRAF packages  IMRED:CDDRED
and  TWODSPEC:APEXTRACT. 

\subsection{Adaptive Optics Imaging}

High spatial resolution  near-IR observations of our entire sample were obtained 
with the Nasmyth Adaptive  Optics Systems (NAOS) and the CONICA camera at 
the 8.2 m telescope  Yepun, which is  part the European Southern Observatory's 
(ESO) Very Large Telescope (VLT) in Cerro Paranal, Chile.  
The data were acquired in service mode during 
the ESO's observing period 84 (April 1$^{st}$ through September 30$^{th}$,  2009). 

To take advantage of the near-IR brightness of our targets,  we used 
the infrared wavefront sensor  and the N90C10 dichroic to direct 90$\%$ of the 
near-IR light to adaptive optics systems and 10$\%$ of the light to the science 
camera. We used the S13 camera  (13.3 mas/pixel and 14$\times$14$''$ field of 
view) and the Double RdRstRd readout mode. The observations were 
performed through the K$_s$ and J-band filters at  5 dithered positions per 
filter. The total exposures times ranged from 1 to 50 s for  the K$_S$-band observations
and from 2 to 200 s  for the J-band observations, depending on the brightness of the target. 
The data were reduced  using the Jitter software, which is part of ESO's data reduction package 
Eclipse\footnote{\tt http://www.eso.org/projects/aot/eclipse/ }.
In \S~\ref{multi_sec}, we use these Adaptive Optics (AO) data to constraint the 
multiplicity of our targets.

\section{Results}

\subsection{Stellar properties}\label{stellar_pro}

Before discussing the circumstellar properties of our targets, which is the main focus
of our paper,  we investigate their  stellar properties.  In this section, we derive their 
spectral types and identify any background objects that might be contaminating our 
sample of PMS stars with IR excesses. 

\subsubsection{Spectral Types}

We estimate the spectral types of our targets by comparing temperature sensitive
features in our echelle spectra to those in templates from stellar libraries. 
We use the libraries presented by Soubiran et al. (1998)  and Montes  (1998). 
The former has a  spectral resolution of 42,000 and covers the  entire 4500 -- 6800  
\AA~spectral range. The Latter has a  resolution of 12,000 and covers the  4000 -- 9000  
\AA~spectral range with some gaps in the coverage.  Before performing the comparison, 
we  normalize all the spectra and take the template and  target to a common resolution. 

Most of our sources are M-type stars,  for which we assign  spectral types based 
on the strength of the  TiO bands centered around 6300, 6700, 7150, and 
7800 \AA. 
We classify G-K stars based on the ratio of  the V I (at 6199 \AA) to Fe I (6200 \AA) line
(Padgett, 1996) and/or on the strength of the Ca I  ( 6112 A)  and Na I  (5890 and 5896 \AA) 
absorption lines  (Montes et al. 1999;  Wilking et al. 2005).
There is also a B-type star in our sample, which we identify and type by the width of the underlying 
H$\alpha$ absorption line (which is much wider than its emission line) and by the 
strength of the Paschen 16, 15, 14, and 13 lines.  The spectral types so derived are listed 
in Table 2. We estimate the typical uncertainties in our spectral classification to be
1 spectral subclass for M-type stars and 2 spectral subclasses for K and earlier type stars.

\subsubsection{Pre-main-sequence stars identification}\label{pms_id}

Background objects are known to contaminate samples of \emph{Spitzer}-selected 
YSO candidates (Harvey et al. 2007, Oliveira et al. 2009). At low flux levels, background galaxies 
are the main source of contamination. However, the optical and near-IR flux cuts we have 
implemented as part of our sample selection criteria seem to have  very efficiently removed any 
extragalactic sources that could remain in the  \emph{Cores to Disks}
catalog of Ophiuchus YSO  candidates.  
At the bright end of the flux distribution, Asymptotic Giant Branch (AGB) stars 
and classical Be stars are the main source of contamination.  
AGB stars are surrounded by shells of dust  and thus have small, but detectable, 
IR excesses.  The extreme  luminosites ($\sim$10$^4$~L$_{\odot}$) of AGB stars 
imply that they can pass our  optical and near-IR flux cuts  even if they are located 
several kpc away. 
Classical Be stars are surrounded by a  gaseous circumstellar disk that is \emph{not}
related to the primordial accretion stage but to the  rapid rotation of the object 
(Porter $\&$ Rivinius, 2003). Classical Be stars exhibit both IR excess (from free-free 
emission) and H$\alpha$ emission (from the recombination of the ionized hydrogen in 
the disk) and thus can easily be confused with Herbig Ae/Be stars (early-type 
PMS stars).  

There is only one B-type star in our target list, object  $\#$ 2. This object is located close 
to the eastern edge of the \emph{Cores to Disks}  maps of Ophiuchus,
shows very little 24 $\mu$m excess, and is not detected at 70 $\mu$m.  
Its weak  IR excess is more consistent with the free-free emission of a classical  Be star than 
with the thermal IR excess produced by circumstellar dust around a Herbig Ae/Be star. 
Since there is no clear evidence that object $\#$ 2 is in fact a pre-main-sequence star, 
we do not include it in our sample of transition disks.  

G and  later type PMS stars can be distinguished from AGB stars by the presence 
of emission lines associated with chromospheric activity and/or accretion.  H$\alpha$ 
(6562  \AA) and the Ca II infrared triplet  (8498, 8542, 8662 \AA) are the most conspicuous 
of such lines.  The Li I  6707 \AA~absorption  line is also a very good indicator of stellar youth 
in mid-K to M-type stars because Li is burned very efficiently in the convective interiors of low-mass 
stars and is depleted soon after these objects arrive on the main-sequence. 
In Table 2, we tabulate the velocity dispersion of the H$\alpha$ emission lines,
the equivalent widths of the  detected Li I  6707 \AA~absorption  lines,  and the identification 
of the Ca II infrared triplets. 
After removing the Be star from consideration, we detect H$\alpha$ emission in 
25 of our targets. All of them exhibit  Li I absorption and/or Ca II emission 
and can therefore be considered bona fide PMS stars.  Six of the targets
exhibit H$\alpha$ absorption.  All of them are M-type stars with small 24 $\mu$m
excesses and no evidence for Li~I absorption nor Ca II emission.  
Since these targets are most likely background AGB stars, we do not include them 
in our sample of transition disks.  Two of our targets, objects \# 12 and 22, show
no evidence for H$\alpha$ emission or absorption. Object \#12 is the well studied 
K-type PMS star Do Ar 21, and we thus include it in our transition disks sample.
Object \#  22 is a M6 star with small 24 $\mu$m excesses and no evidence for Li I 
absorption nor Ca II emission and is also a likely AGB star. 

To sum up, after removing one classical Be star (object \# 2) and seven likely AGB stars 
(objects \# 4, 6, 8, 10, 22, 33, 34), we are left with 26 bona-fide PMS stars.
These 26 objects constitute our sample of transition disks.  As shown in 
Figure~\ref{sample_sel}, all the contaminating objects have  [3.6]-[24] $\lesssim$ 2.5,  
while 24 of the 26 PMS stars have  [3.6]-[24] $\gtrsim$ 2.5.  
This result underscores the need for spectroscopic diagnostics to firmly establish the 
PMS sequence status of  \emph{Spitzer}-selected YSO candidates, especially those
with small IR excesses. 

\subsection{Disk Masses}\label{disk_mass_sec}
 
As shown by Andrews $\&$ Williams (2005,2007),  disk masses obtained from modeling the 
IR and (sub)-mm SEDs of circumstellar disks are well described by a simple linear  
relation between (sub)-mm flux and disk mass. 
From the ratios of model-derived  disk masses to observed mm fluxes  presented by 
Andrews $\&$ Williams (2005) for 33 Taurus stars,  Cieza et al.  (2008) obtained the 
following relation, which we adopt to estimate  the disk masses of our transition disks:
 
\begin{equation}
 M_{DISK}=1.7\times10^{-1}  [(\frac{F_\nu(1.3mm)}{mJy})\times(\frac{d}{140  pc})^2] M_{JUP} 
 \label{eq_mass}
\end{equation}

Based on the standard deviation  in the ratios of the model-derived  masses to observed mm fluxes, 
the above relation gives disk masses  that are within a factor of $\sim$2 of model-derived values; nevertheless,  
larger \emph{systematic} errors can not be ruled out (see Andrews $\&$ Williams, 2007). 
In particular, the models from Andrews $\&$ Williams (2005, 2007)
assume an opacity as a function of frequency of the form $K_{\nu}$ 
$\propto$ $\nu$  and a normalization of $K_0$ =  0.1 gr/cm$^2$ at 1000 GHz. 
This opacity implicitly assumes a gas to dust mass ratio of 100. 
Both the opacity function and the gas to dust mass ratio are 
uncertain and expected to change due to disk evolution processes 
such as grain growth and photoevaporation.  Detailed modeling and a
additional observational  constraints on the grain size distributions and the gas 
to dust ratios will be needed to derive more accurate disk masses for each 
individual transition disk.

Since in the (sub)mm regime disk fluxes behave as $F_{\nu}  \propto \nu^{2\pm0.5}$ (Andrews $\&$ Williams,  2005),
our targets are expected to be brighter (by a factor of $\sim$1.4) at  1.1 mm  than they are at 1.3 mm.
We thus modify Equation~\ref{eq_mass}  accordingly to derive upper limits for the disk masses of the 
objects observed at 1.1 mm  with the CSO.  
The disk masses (and 3-$\sigma$ upper limits) for our sample, derived adopting a distance to Ophiuchus of  
125 pc  (Loinard et al. 2008),  are listed in Table 3.
For two objects, \#12 and  27, we adopt the  disk mass upper limits from Cieza et al. (2008), which
were derived from 850 $\mu$m observations. 
The vast majority of our transition disks have estimated disk masses lower than $\sim$2.0 M$_{JUP}$.
However, two of  them have disk masses typical  of CTTSs  ($\sim$3-5 M$_{JUP}$) and two others have 
significantly more  massive disks ($\sim$10-40 M$_{JUP}$). 

\subsection{Accretion Rates}\label{Acc}

Most PMS stars show H$\alpha$ emission, either from chromospheric activity or 
accretion.  Non-accreting objects show narrow ($<$ 200 km/s) and  symmetric 
line profiles of chromspheric origin, while accreting objects present broad ($>$ 300 km/s) 
and asymmetric profiles produced by large-velocity magnetospheric  accretion columns.
The  velocity dispersions (at 10 $\%$ of the peak intensity), $\Delta V$,  of  the H$\alpha$ emission 
lines of our  transition disks are listed in Table 2.
The boundary between accreting and non-accreting objects has been empirically placed by
different studies at $\Delta V$ between 200 km/s (Jayawardana et al. 2003) and 270 km/s 
(White $\&$ Basri, 2003). 
Since only one  of our objects, source \#  25,  has $\Delta V$  in the 200-300 km/s range,  most accreting and 
non-accreting objects are clearly separated in our sample. Source \# 25 has $\Delta V$  
$\sim$230 km/s and a very noisy spectrum. We consider it to be non-accreting because it has
a very low fractional disk luminosity (L$_D$/L$_{*}$ $<$ 10$^{-3}$, see \S~\ref{sed_mor}).

\begin{figure}
\includegraphics[width=6.0in]{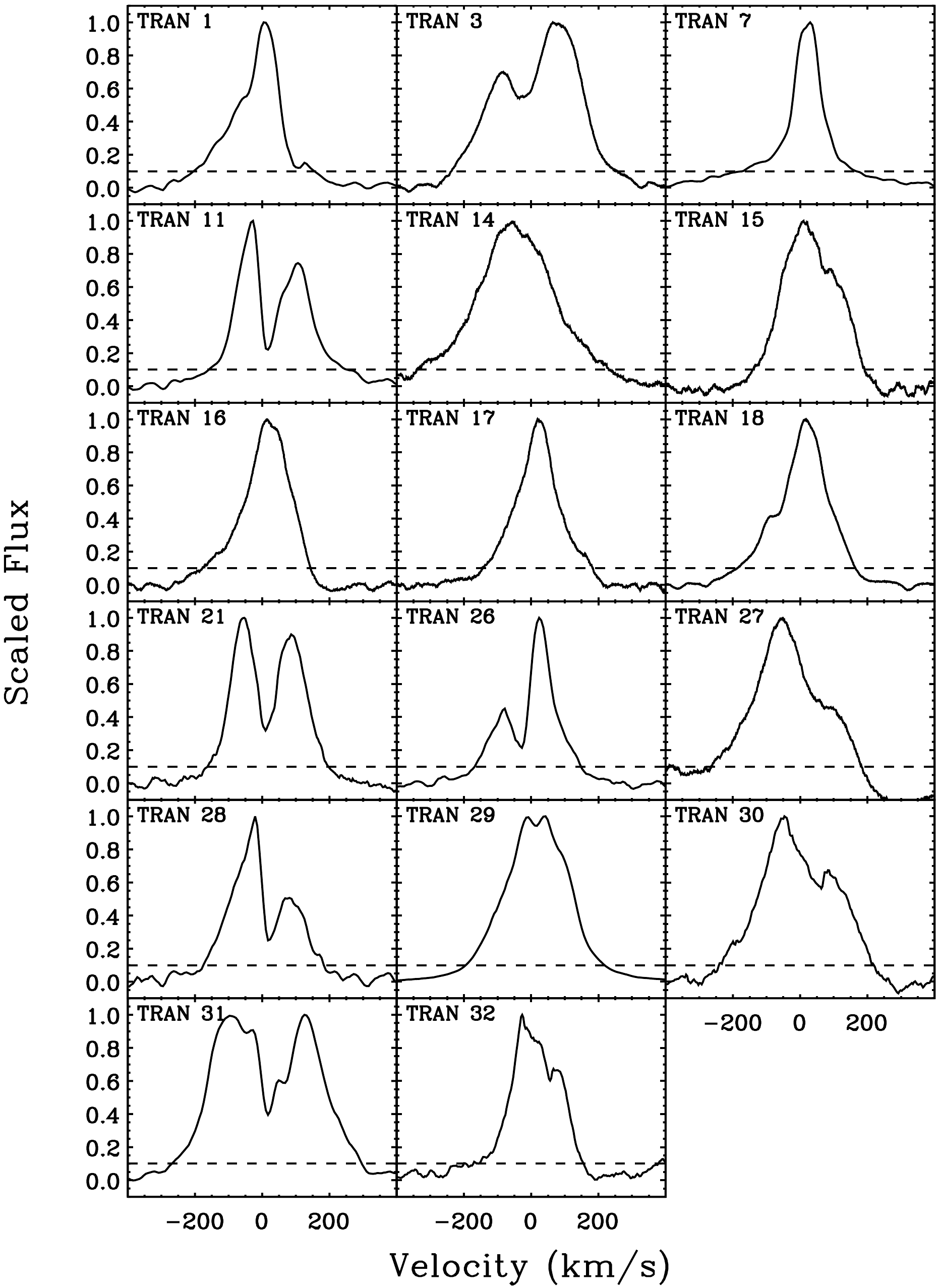}
\caption{The H$\alpha$ velocity profiles of the 17  accreting 
objects in our sample.  The dashed line indicates the 10$\%$ peak intensity,
where $\Delta$V is measured.
They all have $\Delta$V $>$ 300 km/s.
}
\label{prof_acc}
\end{figure}

\begin{figure}[th]
\includegraphics[width=4.0in]{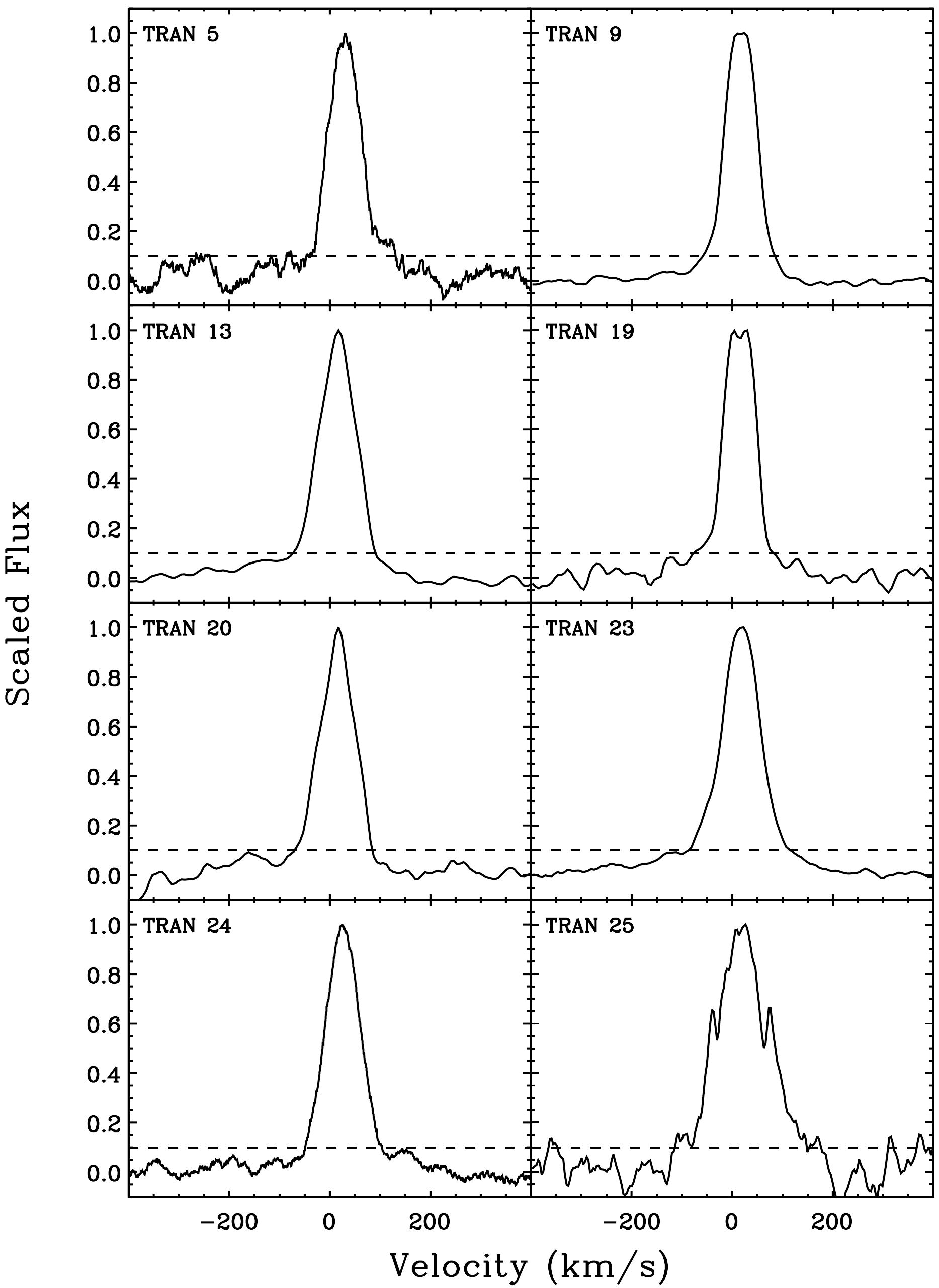}
\caption{
The H$\alpha$ velocity profiles of the 8 non-accreting 
objects in our sample where H$\alpha$ was detected.
The dashed line indicates the 10$\%$ peak intensity,
where $\Delta$V is measured.
Non-accreting objects  show symmetric and  narrow  
($\Delta$V $\lesssim$ 230 km/s)
H$\alpha$ emission, consistent with chromospheric
activity.  
}
\label{prof_non_acc}
\end{figure}

The continuum-subtracted H$\alpha$ profiles for all  the 17 accreting transition disks are shown in 
Figure~\ref{prof_acc}, while those for the 8 non-accreting disks where the H$\alpha$ line
was detected are shown in Figure~\ref{prof_non_acc}.
For accreting objects, the velocity dispersion of the H$\alpha$ line
correlates well with accretion rates derived from detailed models of the 
magnetospheric accretion process. We therefore estimate the accretion rates of our targets
from the width of the H$\alpha$ line measured at 10$\%$   
of its peak intensity, adopting the relation given by Natta et al.  (2004):

\begin{equation}
Log (M_{acc}(M_{\odot}/yr)) = -12.89(\pm0.3) + 9.7(\pm0.7)\times10^{-3} \Delta V (km/s)
\end{equation} 

This relation is valid for 600 km/s $>$ $\Delta V$ $>$  200 km/s  (corresponding to  
10$^{-7}$ M$_{\odot}$/yr $>$ M$_{acc}$  10$^{-11}$ M$_{\odot}$/yr) and can be applied 
to objects with a range of stellar (and sub-stellar)  masses.  
%
%
The broadening of the the H${\alpha}$ line is 1 to 2 orders of
 magnitude more sensitive to the presence of low accretion rates than 
 other accretion indicators such as U-band excess and continuum veiling
 measurements (Muzerolle et al. 2003, Sicilia-Aguilar et al. (2006b),
 and is thus particularly useful to distinguish weakly accreting from
 non-accreting objects. However, as discussed by Nguyen et al. (2009),
 the 10$\%$ width measurements are also dependent on the line profile,
 rendering the 10$\%$ H$\alpha$ velocity width a relatively poor
 \emph{quantitive} accretion indicator, specially at high accretion 
 rates (Muzerolle et al. 1998).

For the 9 objects we consider to be non-accreting, 
we adopt a mass accretion  upper limit of 10$^{-11}$M$_{\odot}$/yr, corresponding to $\Delta V$ = 200 km/s.  
The  so derived  accretion rates (and upper limits) for our sample of transition disks are listed in Table 3.
Given the large uncertainties associated with Equation 2 and the intrinsic 
variability of accretion in PMS stars, these accretion rates should be 
considered order-of-magnitude estimates.

\subsection{Stellar companions}\label{multi_sec}

\begin{figure}[t]
\includegraphics[width=6.5in]{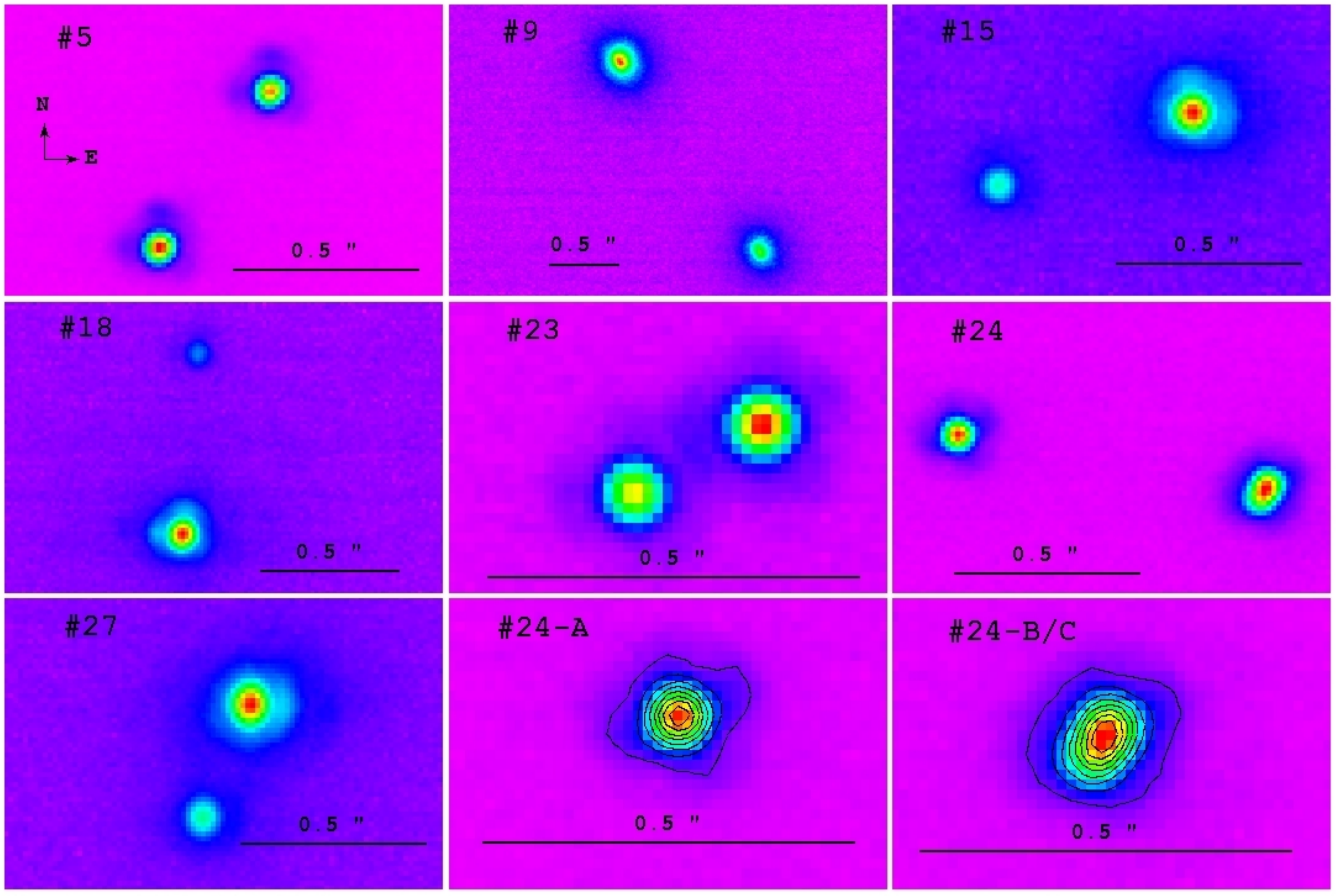}
\caption{
The K-band images of the seven multiple systems that have been detected with our VLT-AO observations.
Target \#24 is a triple system.  The tighter  components  (\#24-B and \#24-C) are not fully resolved, but 
they presence can be inferred from the highly elongated image less than 1$''$ away  from the perfectly round 
PSF of the source \#24-A 
}
\label{multi}
\end{figure}

Seven binary systems were identified in our sample by visual inspection of the VLT-AO images:  targets  \#5, 9, 
15, 18, 23, 24,  and  27 (Figure~\ref{multi}).  The separation and flux ratios of these systems range from 
0.19$''$ to 1.68$''$ and 1.0 to 12, respectively.  All these systems were well resolved in both our J- and K$_{S}$-band images, which 
have typical FWHM values of 0.06$''$-0.07$''$. Targets \#9, 15, 18, and 27 are previously known binaries
(Ratzka et al. 2005; Close et al. 2007), while targets  \#5, 23, and 24  are newly identified multiple systems. 
For each one of these binary systems,  we searched for additional tight  companions by comparing each other's PSFs.  
The PSF pairs were virtually identical  in all cases, except for  target \#24.  The north-west component of  this
target has  a perfectly round PSF, while the south-east component,  0.84$''$ away,  is clearly elongated.
Since variations in the PSF shape are not expected within such small angular distances and this behavior is seen in both the 
J- and K$_{s}$-band images, we conclude that target \#24 is a triple system.  We use the PSF of the wide component 
(object \#24-A) to  model the  image of the tight components (object \#24-B/C).  We find that the image of the tight components 
 is well reproduced by two equal-brightness objects 4 pixels apart (corresponding to $\sim$0.05$''$ or 6.6 AU) at a 
position angle of $\sim$30 deg.

We have also searched in the literature for additional companions in our sample that 
our VLT observations could have missed.  In addition to the seven multiple systems discussed 
above, we find that source \# 12 (Do Ar 21) is a binary system with a projected separation of just 0.005$"$  (corresponding to 0.62 AU), 
recently identified by the Very Long Baseline Array (Loinard et al. 2008).  We also find that source \# 27 is in fact a triple system. 
The ``primary" star in the VLT observations is itself  a spectroscopic binary with a 35.9 d period  and an estimated separation 
of 0.27 AU (Mathieu et al. 1994).
Additional multiplicity constraints exists for 2 other of our targets. 
Target \#21 has been observed with the lunar occultation technique
 (Simon et al. 1995) and a  similar-brightess companion can be ruled out down to $\sim$1 AU. 
 Similarly,  target \#13 has been found to have a constant radial velocity ( $\sigma_{vel} <$ 0.25 km/s)
during a 3 yr observing campaign (Prato,  2007).

For the apparently single stars, we estimated the detection limits at 0.1$''$  separation from the 5-$\sigma$ noise 
of PSF-subtracted images. For these objects, however, there are no comparison PSF's  available to
perform the subtraction.  Instead, we subtract a PSF constructed by azimuthally smoothing the  image 
of the target itself,  as follows. For each pixel in the image, the separation from the target's centroid is calculated, 
with sub-pixel accuracy. The median intensity at that separation, but within an arc  of  30 pixels in length, 
is then subtracted from the target pixel.  Thus, any large scale, radially symmetric  structures are removed.
The detection limits at 0.1$''$  separation, obtained as described above,  
range from  2.4 to 3.5 mag, with a median value of 3.1 mag (corresponding to a flux ratio of  17).
The separations, positions angles, and flux ratios of the multiple systems  in our sample are shown in Table 2, 
together with the flux ratio limits of companions for the targets that appear to be single.

\section{Discussion}

\subsection{Sample properties}

In this section, we investigate the properties of transition disks as  a group of objects.  
We discuss the properties of our targets  (disk masses, accretion 
rates,  multiplicity,  SED morphologies, and fractional disk luminosities)
and compare them to those of non-transition disks  in order
to place our sample into a broader context of disk evolution. 

\subsubsection{Disk Masses and Accretion Rates}\label{m_a_sec}

To compare the disk masses and accretion rates of our sample 
of transition disks to those of non-transition disks, we collected such
data from the literature for Ophichus objects with   \emph{Spitzer} colors [3.6]-[4.5]  $>$  0.25 
and [3.6]-[24] $>$ 1.5.   These are objects that have both  near-IR and mid-IR excesses 
and therefore are likely to have ``full disks", extending inward to the dust sublimation 
radius. 
We collected disk masses for non-transition disks from Andrews $\&$ Williams (2007).
This comparison sample is appropriate because our mm~flux to disk~mass relation 
(i.e., Equation~\ref{eq_mass})
has been calibrated using their disk models. 
The accretion rates for our comparison sample come from Natta et.  (2006),
and have been calculated from the luminosity of the Pa$\beta$ line.
This comparison sample is also appropriate because the relations between 
the luminosity of  the Pa$\beta$ line and accretion rate and between the H$\alpha$ full 
width at 10$\%$ intensity  and accretion rate (which we use for our transition disk sample)  
have both been calibrated by Natta et al. (2004) using the same detailed models of magnethospheric 
accretion. 

\begin{figure}[htp]
\includegraphics[width=3.3in]{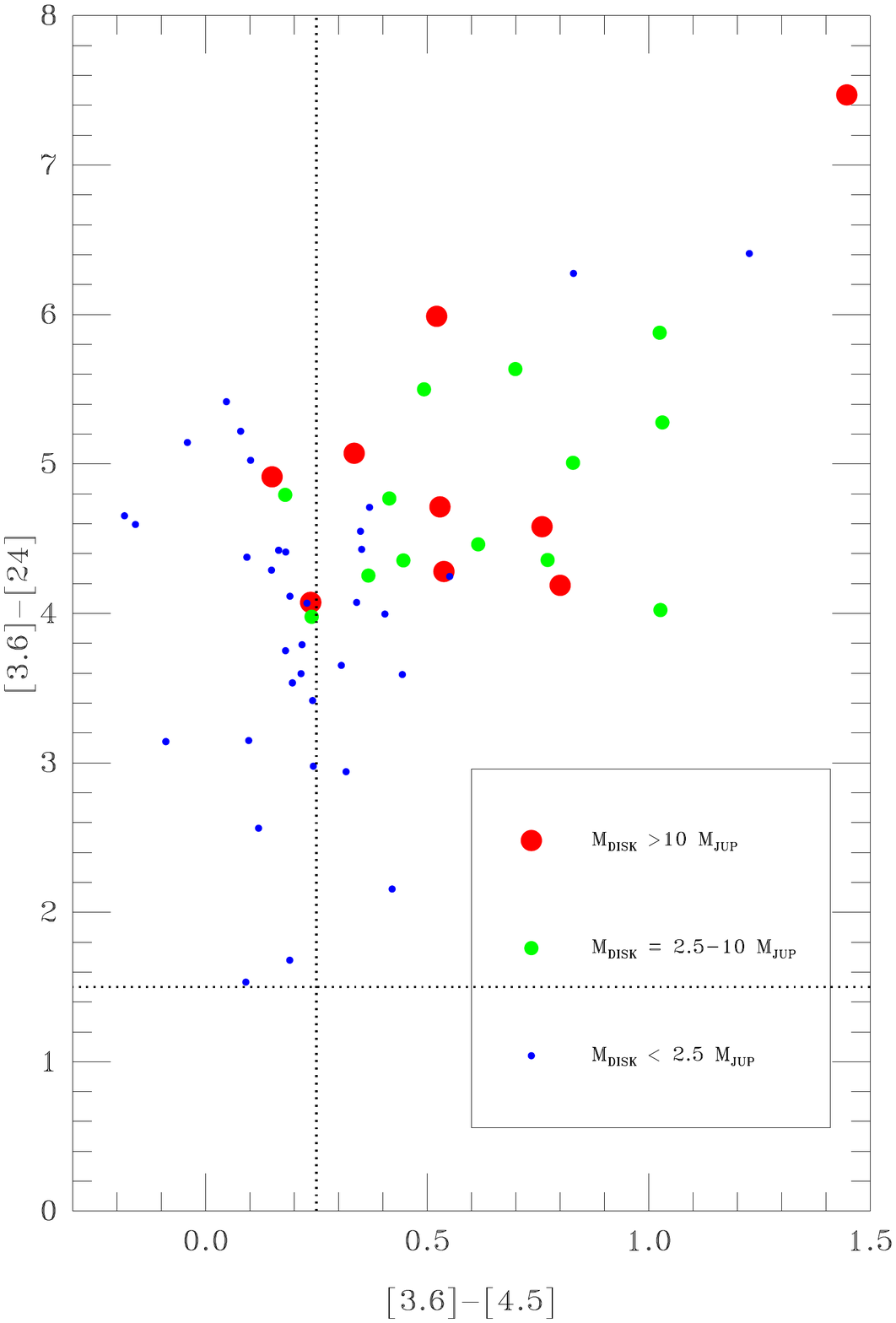}
\includegraphics[width=3.3in]{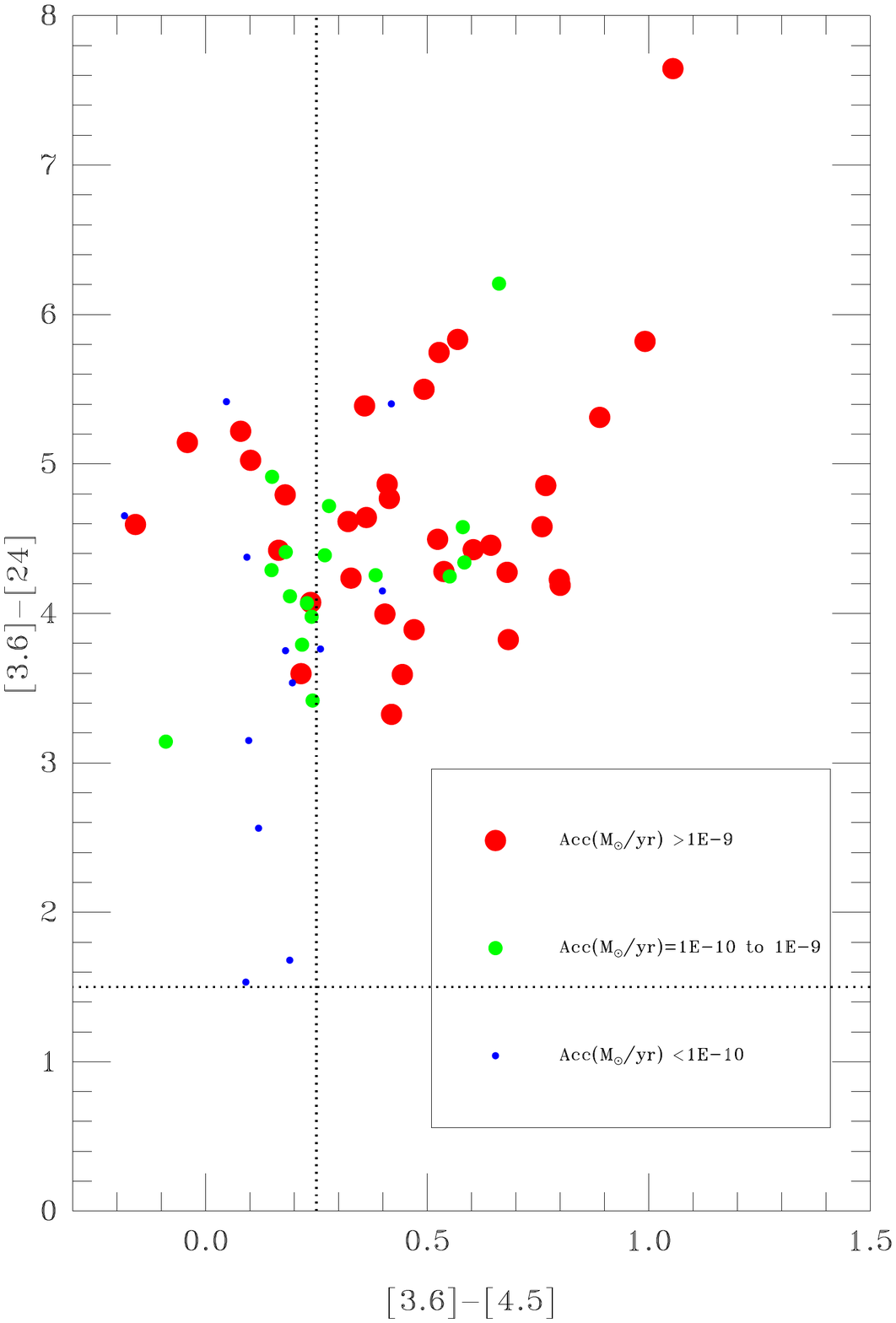}
\caption{
The [3.6] -[24] vs [3.6]-[4.5] colors of  Ophiuchus PMS stars with disks of different masses (left) 
and accretion rates (right). Transition disks (in the left side of the figure, with [3.6]-[4.5] $<$ 0.25 ) 
tend to have much lower disk masses and accretion rates  than non-transition objects.  
The  disk masses of non-transition disk sample have been taken from Andrews $\&$  Williams (2007).
The  accretion rates of non-transition disks have  been taken from  Natta et al.  (2006).
}
\label{m_a_t_nt}
\end{figure}

\begin{figure}[t]
\includegraphics[width=6.0in]{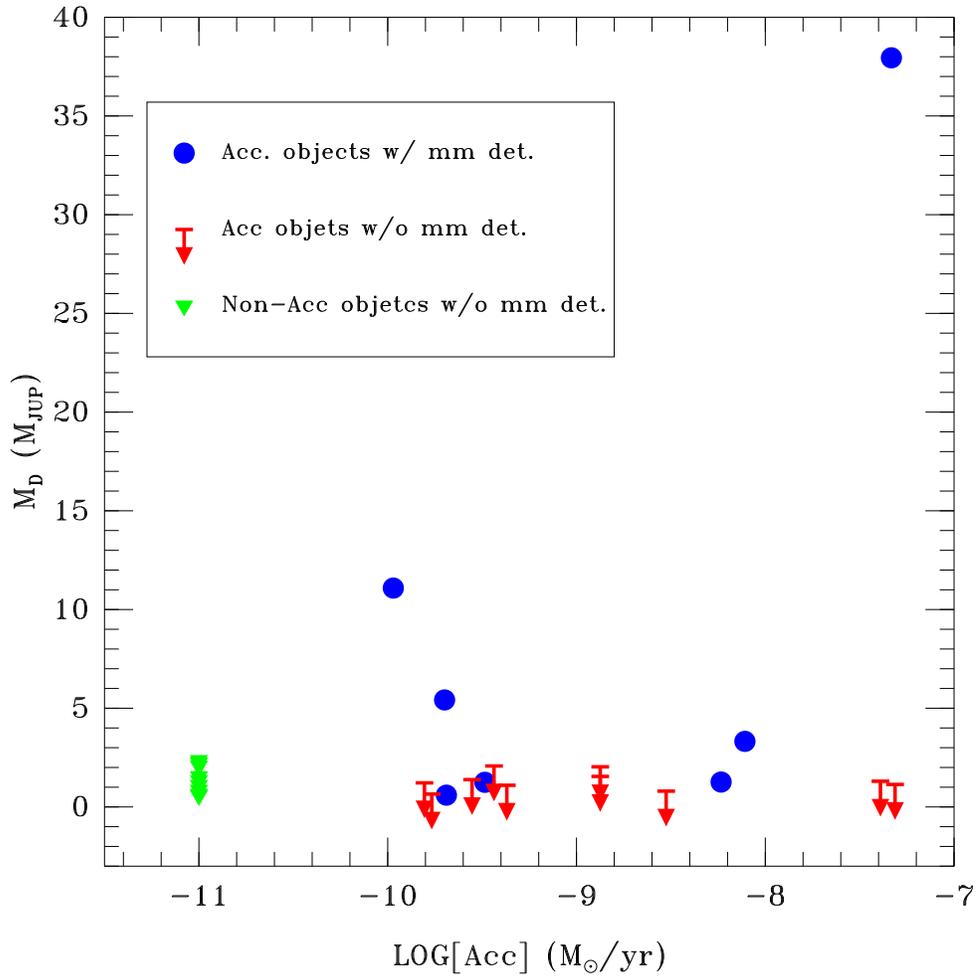}
\caption{
Disk masses vs. accretion rates  for accreting objects with mm detections (blue circles),
accreting objects with mm upper  limits   (red arrows), and non-accreting objects with
mm upper limits (green triangles). None of the non-accreting objects was detected at mm wavelength.}
\label{mass_vs_acc}
\end{figure}

Figure~\ref{m_a_t_nt} shows that transition disks tend  to have much lower disk masses 
and accretion rates than non-transition objects.  
We also find a strong connection between the magnitude of the 
mid-IR excess and both disk mass and accretion rate: objects with little ($\lesssim$ 4  mag)
24 $\mu$m excess tend to have small disk masses and low accretion rates. 
Among our transition disk sample,  we also note that all of the mm detections 
correspond to accreting objects and  that even some of the strongest accretors 
have  very low disk masses (see Figure~\ref{mass_vs_acc}). 
In other words, massive disks are the most likely to accrete, but  some 
low-mass disks can also be strong accretors. 

These results can help reconcile the apparently contradictory findings of previous studies of 
transition disks. As discussed in \S~\ref{intro},  Najita et al. (2007) studied a sample of 12
transition objects in Taurus and found that they have  a median disk mass of 25 M$_{JUP}$,
which is  $\sim$4 times larger than the rest of the disks in Taurus, while Cieza et al. (2008)
found that the vast majority of their 26  transition disk targets had very small ($<$ 2 M$_{JUP}$) disk
masses. 
Our results now show that transition disks are a highly  heterogeneous group of objects, whose
``mean properties" are highly dependent on the details of  the sample selection criteria.
On the one hand,  the sample studied by Cieza et al.  (2008) was dominated by weak-line T Tauri 
stars. This explains the low disk masses they found, as  lack of accretion is a particularly  good indicator 
of a low disk mass (See Figure~\ref{mass_vs_acc}).
One the other hand, Najita et al. (2006) drew their sample from the \emph{Spitzer}
spectroscopic survey of Taurus presented by Furlan et al. (2006), which in turn selected their sample
based on mid-IR  colors and fluxes from the Infrared Astronomical Satellite (IRAS). 
As a result, the \emph{Spitzer} spectroscopic survey of Taurus was clearly biased towards the brightest objects in
the mid-IR.  This helps to explain the high disk masses obtained for the transition disks studied by Najita et al. (2006), 
as objects with strong mid-IR excesses tend to have higher disk masses (Figure~\ref{m_a_t_nt}).

\subsubsection{Multiplicity}\label{multi_text}

As discussed in \S~\ref{intro}, dynamical interactions between the disk and stellar companion is one of the 
main mechanisms proposed to account for the opacity holes of transition disks. 
However, the fraction of transition disks that are in fact close binaries still remains to be established. 
If binaries play a dominant role in transition disks, one would expect a higher incidence of close binaries in transition
disks that in non-transition disks.  
In order to test this hypothesis, we compare the distribution of binary separations of our sample 
of transition disk to that of non-transition disk (and disk-less PMS stars).  
We drew our sample of non-transition disks and disk-less PMS stars from the compilation
of Ophiuchus binaries presented by Cieza et al. (2009). 

\begin{figure}[t]
\includegraphics[width=6.0in]{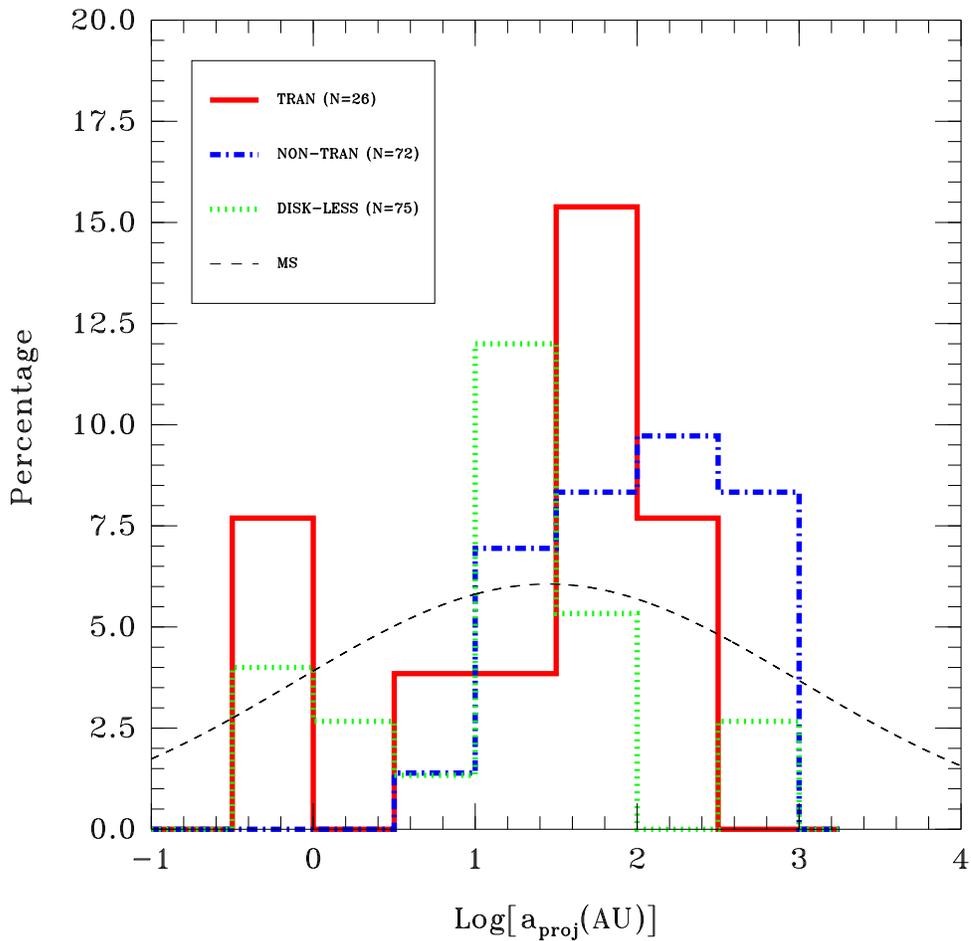}
\caption{Distribution of  companion separations for our transition disk sample (red dashed line), non-transition
disks (blue dash-doted line), and disk-less stars (green dotted line). Spectroscopic binaries 
have been assigned a projected separation of 0.5 AU. The total number of objects (i.e., single + multiple systems) 
in each sample are shown in parenthesis.  The distribution of binary separations for main sequence (MS) 
solar-type stars (Duquennoy $\&$ Mayor, 1991) is shown for comparison.
}
\label{multi_fig}
\end{figure}

As shown by Figure~\ref{multi_fig},  disk-less PMS stars tend to have companions 
at smaller separations than stars with regular, non-transition disks. 
According to a two-sided Kolmogoro--Smirnov (KS) test, there is only a  0.5$\%$ probability that the distributions of binary separations of 
non-transition disks and disk-less stars have been drawn from the same parent population.
As discussed by Cieza et al. (2009), this result can be understood in terms of the effect binaries have 
on circumstellar disks lifetimes via the tidal truncation of the \emph{outer} disk.
For instance, a binary system with a 30 AU separation is expected to initially have individual disks that are 
$\sim$10-15 AU in radius. Given that the viscous timescale is roughly proportional to the size of the disk, such small
disks are likely to have very small accretion lifetimes,  smaller than the age of the sample.  

For systems with binary separations much smaller than the size of a typical disk, the outer 
disk can survive in the form of a circumbinary disk, with a tidally truncated inner hole 
with a radius $\sim$2$\times$  the orbital separation (Artymowicz $\&$ Lubow, 1994).
Such systems could be classified as transition disks based on their SEDs.
We do not see an increased incidence of binaries with separations in the 
$\sim$8-20 AU  range, the range where we are sensitive to companions
 that could in principle carve the inferred inner  holes.
In fact,  the incidence of binaries in this separation range is lower for our transition 
disks than it is  for the samples of both non-transition disks and disk-less stars. 
This suggests that stellar companions at 8-20 AU separations are \emph{not} 
responsible for a large fraction of the transition disk  population.  
Given the size of our transition disk sample, however, 
this result carries  little statistical weight, and should be considered only a hint. 
According to a KS test, the distribution of  binaries separations of transition objects 
is indistinguishable from that of both  non-transition disks  (p = 28$\%$) and disk-less 
stars (p  = 12$\%$).
On the other hand,  the fact that 2 of our transition disk targets 
have previously known companions at sub-AU separations suggests that circumbinary disks
around  very tight systems (e.g., r $\lesssim$ 1-5 AU) could indeed represent an important component of
 the transition disk population. Unfortunately, the presence of  tight 
 companions (r $\lesssim$ 8 AU) remains completely unconstrained for most of our  sample. 
A complementary radial velocity survey to find the tightest companions 
is highly desirable to firmly establish the fraction of  transition disks that 
could be accounted for by close binaries.

\subsubsection{SED morphologies and fractional disk luminosities}\label{sed_mor}

In addition to the disk mass, accretion rate, and multiplicity,  the SED morphology and fractional disk
luminosity of a transition disk can provide important clues on  the nature of the object.
In this section, we quantify the SED morphologies and fractional disk luminosities of our transition
objects in order to compare them to the values of objects with ``full disks".

The diversity of SED morphologies seen in transition disks can not be properly captured 
by the traditional classification scheme of young stellar objects (i.e., the class I through III system), 
which is based on the slope of the SED between 2 and 25 $\mu$m (Lada 1987).
Instead, we quantify  the  SED  ``shape" of our targets adopting the two-parameter scheme 
introduced by Cieza et al. (2007), which is based on the longest wavelength at which the 
observed flux is dominated by the stellar photosphere, $\lambda_{turn-off}$, and the slope 
of the IR excess,  $\alpha_{excess}$, computed as  dlog($\lambda$F)/dlog($\lambda$) between 
$\lambda_{turn-off}$ and 24 $\mu$m.  
The $\lambda_{turn-off}$ and  $\alpha_{excess}$ values for our entire sample 
are listed in Table 3. 

The $\lambda_{turn-off}$ and  $\alpha_{excess}$ parameters provide useful 
information on the structure of the disk.  In particular,  $\lambda_{turn-off}$ 
clearly correlates with the size of the inner hole as it depends on the temperature of the 
dust closest to the star.  Wien's displacement law implies that \emph{Spitzer}'s
3.6, 4.5, 5.8, 8.0, and 24 $\mu$m bands best probe dust temperatures of roughly 
780K, 620K, 480K, and 350K, and 120K, respectively. For stars of solar luminosity, 
these temperatures correspond  to circumstellar distances of  $\lesssim$1 AU for the IRAC 
bands and $\sim$10 AU for the 24 $\mu$m  MIPS band. 
Similarly,  $\alpha_{excess}$ correlates well with the sharpness
of the hole: a sharp inner hole,  such as that of CoKu Tau/4,  is characterized by a 
large and positive $\alpha_{excess}$ value, while a continuous disk that
has undergone significant grain growth and dust settling  
is characterized by a  large negative $\alpha_{excess}$ value (Dullemond $\&$ Dominik, 2004).
However, the $\lambda_{turn-off}$ and  $\alpha_{excess}$ parameters should be 
interpreted with caution. 
Deriving accurate disk properties from a SED requires detailed modeling because
other properties, such as stellar luminosity and disk inclination, can also affect
the $\lambda_{turn-off}$ and  $\alpha_{excess}$ values.

The fractional disk luminosity, the ratio of the disk luminosity to the
stellar luminosity (L$_{D}$/L$_{*}$), is another important quantity that
relates to the evolutionary status of a disk. On the one hand, typical primordial 
disks around CTTSs have L$_{D}$/L$_{*}$ $\sim$ 0.1 as they have 
optically thick disks that intercept (and reemit in the IR) $\sim$10$\%$ of the 
stellar radiation.  On the other hand, debris disks show  L$_{D}$/L$_{*}$ 
values $\lesssim$ 10$^{-3}$  because they have optically thin disks that intercept 
and reprocess $\sim$10$^{-5}$--10$^{-3}$ of the star's light  (e.g., Bryden et al. 2006). 

We estimate L$_{D}$/L$_{*}$  for our sample of  transition disks\footnote{for completeness,
we also estimate analogous L$_{D}$/L$_{*}$  values for the likely Be and AGB stars in our sample.}
using the following  procedure.  First, we estimate  L$_{*}$ as in Cieza et al (2007), 
by integrating over the broad-band colors given by Kenyon $\&$ Hartmann 
(1995) for stars of the same spectral type as each  one of the stars in our sample.
The integrated fluxes were normalized to the J-band magnitudes corrected for extinction,
adopting  A$_{J}$ =1.53 $\times$ E(J-K$_{S}$ ), where E(J-K${_S}$) is the 
observed color excess  with  respect to the expected photospheric color. 
Then, we estimated L$_{D}$ by integrated the estimated disk fluxes, the observed 
fluxes (or upper limits) minus the contribution of  stellar photosphere at  each wavelength 
between $\lambda_{turn-off}$ and 1.3 mm (or 850 $\mu$m for  targets \#12 and 27).
The disk was assumed to contribute 30$\%$ of the flux at  $\lambda_{turn-off}$ 
(e.g.,  a small but non-negligible amount,  consistent with our definition of  
$\lambda_{turn-off}$).

The near and mid-IR luminosities of our transition disks are well constrained because their SEDs are  
relatively well sampled at  these wavelengths. However, their far-IR  luminosities remain
more poorly constrained.  Only 9 of our 34  targets have ( 5-$\sigma$ or better)  detections  
at 70 $\mu$m listed in the \emph{Cores to Disks}  catalog.  
For the rest of the objects, we have obtained 5-$\sigma$ upper limits, from the 
noise of the 70 $\mu$m images at the location of the targets, in order to fill the gap in their
SEDs between 24 $\mu$m and the millimeter. 
In particular, we used 120$''$ $\times$ 120$''$ cut  outs\footnote{available at \tt http://irsa.ipac.caltech.edu/data/SPITZER/C2D/index\_cutouts.html}
of the Ophichus mosaic created by the \emph{Cores to Disks} project  from the ``filtered'' Basic Calibrated Data.
We calculated the noise as 1-$\sigma$~=~2$\times$1.7(rms$_{sky}$)$\sqrt{N}$, where rms$_{sky}$
is the flux rms (in mJy) of an annulus centered on the source with an inner and outer radius of 40 
and 64$''$, respectively.  N is the number of (4$''$) pixels in a circular aperture of 16$''$ in radius, and 1.7 is the 
aperture correction appropriate for such sky annulus and aperture size \footnote{http://ssc.spitzer.caltech.edu/mips/apercorr/}.
The factor of 2 accounts for the fact that the images we use  have been resampled to  pixels that are half the linear 
size of the original pixels, artificially reducing the noise of the  images (by $\sqrt{       \frac{N_{res}}{N_{orig}}    }$,
where N$_{res}$  and N$_{orig}$ are, respectively,  the number of resampled and original pixels in the sky annulus).  

The LOG(L$_D$/L$_*$) values for our transition disk sample, ranging from -4.1 to -1.5 
are listed in Table 3. As most of the luminosity of a disk extending inward to the 
dust sublimation temperature is emitted in the near-IR, L$_D$/L$_{*}$ is a very strong function 
of $\lambda_{turn-off}$: the shorter the $\lambda_{turn-off}$ wavelength, the higher the 
fractional disk luminosity.  For objects with $\lambda_{turn-off}$ $<$ 8.0 $\mu$m,  
the 70 $\mu$m flux represent only a minor contribution to the total disk luminosity. 
On the other hand, objects with with  $\lambda_{turn-off}$ $=$ 8.0 $\mu$m (i.e., objects with 
significant excess at 24 $\mu$m only) have much lower L$_D$/L$_*$ values, and the 70 $\mu$m emission 
becomes a much larger fraction of the total disk luminosity.  As a result, the L$_D$/L$_*$ values of objets 
with $\lambda_{turn-off}$ $=$ 8.0 $\mu$m and no 70 $\mu$m detections should be considered upper limits.

\begin{figure}[t]
\includegraphics[width=6.0in]{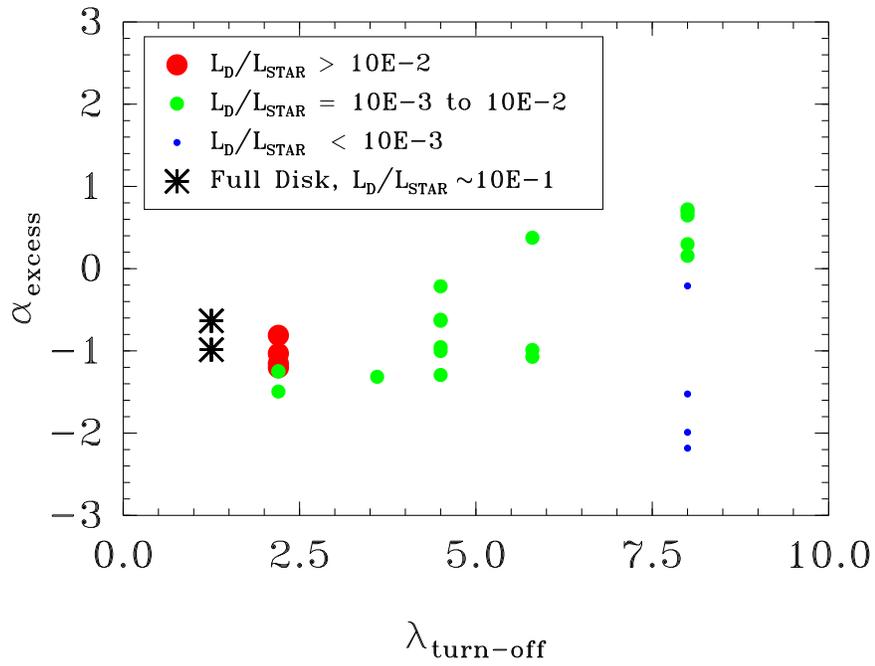}
\caption{$\alpha_{excess}$ vs. $\lambda_{turn-off}$ for  objects with different L$_{D}$/L$_{*}$ values.
The loci of  ``typical" CTTS  with full disks are shown for comparison. Transition disks occupy a much larger parameter 
space}
\label{LTO_FDL}
\end{figure}

As shown in Figure~\ref{LTO_FDL}, our transition disks span the range  
$\lambda_{turn-off}$ = 2.2 to 8.0 $\micron$ and $\alpha_{excess}$ $\sim$ -2.5 
to 1, while typical CTTSs  occupy a much more restricted region of the parameter 
space. Based on the median and quartile SEDs presented by Furlan et al (2006), 
 50$\%$ of late-type (K5--M2) CTTSs have $\lambda_{turn-off}$ = 1.25 $\micron$, 
 -1.0 $>$ $\alpha_{excess}$  $>$ -0.64, and fractional disk luminosities in 
 the  0.07 to 0.15 range. 
 As discussed by Cieza et al. (2005), 
 CTTS are likely to have J-band excesses. However, since this excess is at  the  
 $\lesssim$ 30-40$\%$ level, the observed J-band fluxes are still dominated by the
 stellar photospheres, satisfying our $\lambda_{turn-off}$ definition. 
 Figure~\ref{LTO_FDL}  also shows that, as expected, transition disks with
 different L$_{D}$/L$_{*}$ values occupy different regions of the  $\alpha_{excess}$ vs.
$\lambda_{turn-off}$ space. Objects with L$_{D}$/L$_{*}$ $>$  10$^{-2}$
lay close to the CTTSs loci, while objects with L$_{D}$/L$_{*}$ $<$  10$^{-3}$ 
all have $\lambda_{turn-off}$ = 8.0 $\micron$  and $\alpha_{excess}$  $\lesssim$ 0.0.
The objects in the latter group are all non-accreting. Their very  low fractional disk luminosities 
and lack of accretion suggest they are already in a debris disk stage
(see \S~\ref{photo_sec}). 

\subsection{The Origin of the Opacity Holes}

The four different  mechanisms that have been proposed to explain the inner 
holes of transition disks can be distinguished when disk masses, accretion rates and multiplicity 
information are available (Najita et al., 2007; Cieza, 2008; Alexander, 2008).  In this section, 
we compare the properties of our transition disk sample to those
expected for objects whose opacity holes have been produced by these four mechanisms: 
photoevaporation, tidal truncation in close binaries, grain-growth, and giant planet formation. 
 
 \subsubsection{Photoevaporation}\label{photo_sec}
 
According to photoevaporation  models  (e.g., Alexander et al. 2006),  extreme UV
photons, originating in the stellar chromosphere,  ionize and heat
the circumstellar hydrogen. Beyond some critical radius, the thermal velocity of the ionized
hydrogen exceeds its escape velocity and the material is lost as a wind.
At early stages in the evolution of the disk, the accretion rate across the disk dominates over the evaporation
rate, and the disk undergoes standard viscous evolution.  Later on, as the accretion
rate drops to the level of the photoevaporation rate, the outer disk is no longer able to resupply the inner
disk with material, the inner  disk drains on a viscous timescale,  and an inner hole
is formed. 
Once an inner hole has formed, the entire disk dissipates very rapidly from the
inside out as a consequence of the direct irradiation of the disk's inner edge.  
Thus, transition disks created by photoevaporation are expected to have relatively low masses  (M$_{DISK}$  $<$  $5 M_{JUP}$) 
and negligible accretion ($<$10$^{-10}$ M$_{\odot}$yr$^{-1}$).

\begin{figure}[htp]
\includegraphics[width=6.0in]{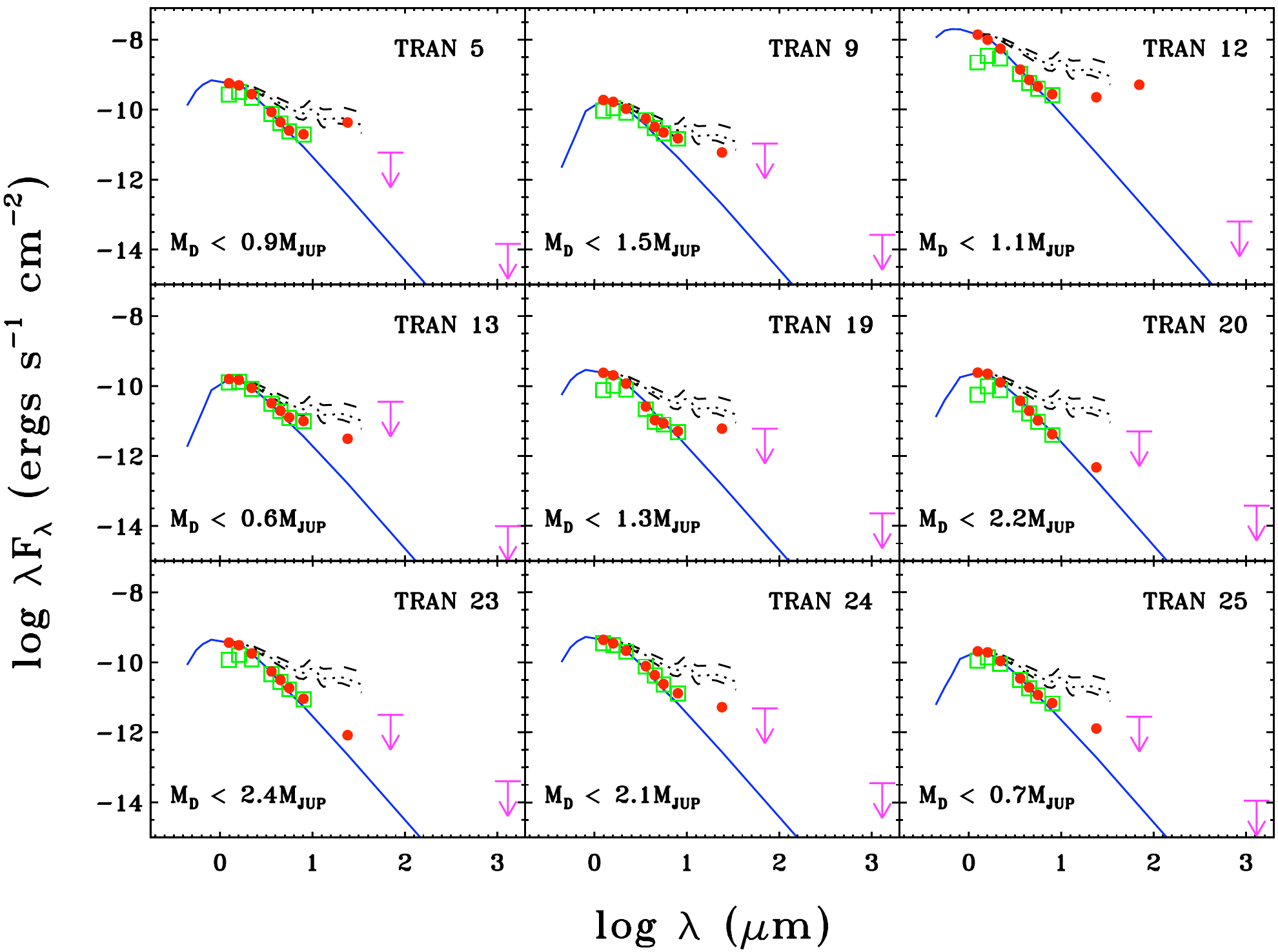}
\caption{The SEDs of  our non-accreting transition disk targets. All of them have low disk masses ($<$ 2.5 M$_{JUP}$),
and  are consistent with photoevaporation.
The  filled circles are detections while the arrows represent 3-$\sigma$ limits. The 
open squares correspond to the observed optical and near-IR fluxes before 
being corrected for extinction using the A$_J$ values listed in Table 3 and the 
extinction curve provided by the Asiago database of photometric systems  
(Fiorucci $\&$ Munari  2003). For each object, the average of the two R-band magnitudes (from the USNO-B1 catalog)
listed in Table 1 has been used.  The solid blue line represent the stellar photosphere normalized to the extinction-corrected
 J-band.  The dotted lines correspond to the median mid-IR SED of K5--M2 CTTSs calculated by Furlan et al. (2006). 
The dashed lines are the quartiles. 
}
\label{sed_non_acc}
\end{figure}

\begin{figure}[htp] 
\includegraphics[width=6.0in]{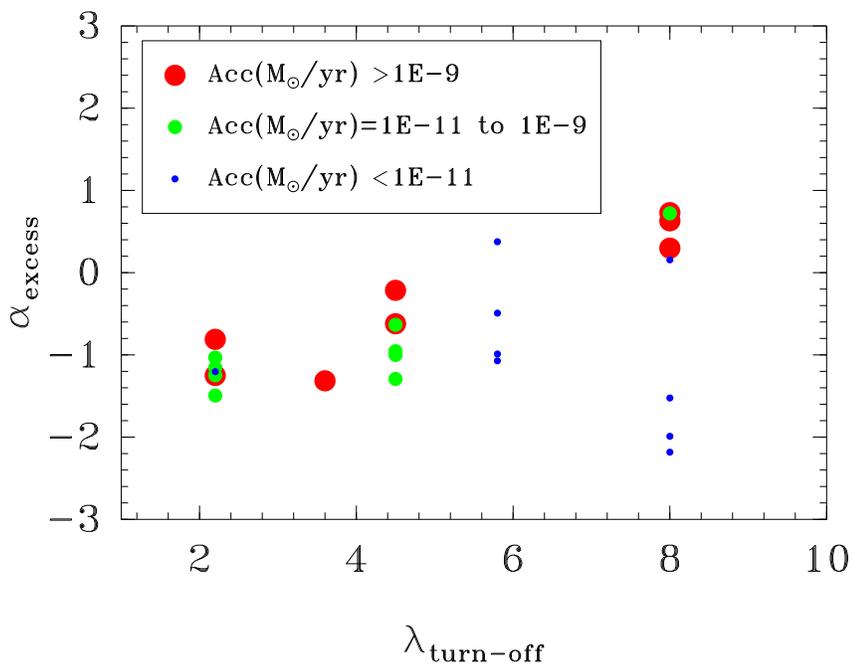}
\caption{The  $\alpha_{excess}$ (the slope of the IR excess) vs.  $\lambda_{turn-off}$ (the wavelength at
which the excess becomes significant) for our sample of transition disks. 
The masses and accretion rates  are indicated by  different symbols.
The 4 accreting (targets \#11, 2, 31, and 32)  and the  2 non-accreting (targets \#5 and 19) 
objects with  $\alpha_{excess} ~\gtrsim$~0  are suggestive of sharp inner holes.}
\label{l_a_m_a}
\end{figure}

We find that 9 of our 26  transition disks have low disk mass ($<$ 0.6-2.5 M$_{JUP}$) and negligible accretion  
(M$_{acc}$ $<$ 10$^{-11}$ M$_{\odot}$/year) and are hence consistent  with  photoevaporation.  
The  SEDs  of these objects are shown in Figure~\ref{sed_non_acc}. 
These  objects appear all  in the lower left corner of Figure~\ref{mass_vs_acc}, which shows the 
disk mass as a function of accretion rate for our sample.
However, they occupy different regions of the $\alpha_{excess}$ vs $\lambda_{turn-off}$ 
parameter space.  The majority of  these objects  have 
 $\alpha_{excess}$  $\lesssim$ -1, the typical value of   CTTS with ``full disks" (see Figure ~\ref{l_a_m_a}). 
 These  objects are therefore consistent with relatively flat, radially continuous disks. 
 Two  objects,  targets \# 5 and 19, have $\alpha_{excess}$  $\gtrsim$ 0, and are thus suggestive of sharp 
 inner holes.  Target \#12 also has an steep SED, but  between 24 and 70 $\mu$m, suggesting 
 a larger inner hole
\footnote{The mid-IR excess of target \#12 (DoAr 21) comes from heated material that is at least 
100 AU away from the star and it has been suggested it may originate from a small-scale 
photodissociation region powered by DoAr 21 itself and not  necessarily from a circumstellar disk 
(Jensen et al.  2009).}.  
Sharp  inner holes could also arise from  dynamical  interactions between the disk 
and large bodies within it. 
 Photoevaporation and dynamical interactions are not mutually exclusive processes,
 as photoevaporation can also operate on a dynamically truncated disk as 
 long as the disk mass and accretion rates are low enough. 
 Coku Tau/4 is perhaps the most  extreme example of this scenario.  It is a 8 AU separation, equal-mass, binary system  
 (Ireland \& Kraus, 2008) resulting on  a $\alpha_{excess}$  value of  $\sim$2.  
Just like targets \#5,  12, and 19, Coku Tau/4 lacks  accretion and has a very low disk mass 
(Najita et al. 2007), and is consequently also susceptible to photoevaporation. 
Our  VLT-AO  observations reveal no stellar companion within $\sim$8 AU of
targets \#5, 12, or 19. However,  target  \#12 is already known to be very tight binary 
(r $\lesssim$ 1 AU), and the presence of  stellar (or planetary mass)
companions remains completely  unconstrained  for targets \#5 and 19. 

Photoevaporation provides a natural mechanism for objects to transition from
the primordial to the debris disk stage.   
Once accretion stops, photoevaporation is capable of removing the remaining gas
in  a very short ($<<1$ Myr) timescale (Alexander et al. 2006). 
As the gas in the disk photoevaporates,  it is likely to carry with it the smallest grains present in the disk.  
What is left represents the initial conditions  of a debris disk: a gas poor disk with large grains,  planetesmals and/or planets. 
The very short photoevaporation  timescale implies that some of the systems shown in Figure~\ref{sed_non_acc} could 
already be in a debris disk stage. 
In fact, the 4  non-accreting objects with L$_{D}$/L$_{*}$ $<$  10$^{-3}$ have properties 
that are indistinguishable from those of  young debris disks like AU Mic and GJ  182 (Liu et al. 2004). 
These  4 objects are \# 12, 20, 23, and 25. 
These objects might have already formed planets or might be in  a multiple system (like target \#12 is). 
Either way, most of the  circumstellar gas and dust has already been depleted, 
and what we are currently observing could be the initial architectures of different debris systems, now 
subjected to processes such as dynamical interactions,  the Poynting--Robertson effect and radiation pressure.
The rest of the non accreting objects, \#5, 9, 13,  19, and  24,  have L$_{D}$/L$_{*}$ 
$>$  10$^{-3}$ and could be in the process of being photoevaporated. 

\subsubsection{Close binaries}
 
Circumbinary disks are expected to be tidally truncated and have inner holes with a radius $\sim$2$\times$ 
the orbital separation (Artymowicz $\&$ Lubow, 1994).  It is believed that most PMS stars are in multiple 
systems (Ratzka et al. 2005) with the same semi-major axis distribution as MS solar-type stars  (e.g., a lognormal 
distribution centered at $\sim$28 AU). As a result, $\sim$30$\%$ of  all Ophiuchus binaries are expected to have
separations in the $\sim$1-20 AU range.  \emph{If the primodial disk has survived in such systems}, it is likely
to be in the form of a circumbinary disk with tidally truncated inner holes up to 40 AU in radius. 
These close binary systems could hence be classified as transition disks based on their IR SEDs.

Four of our targets have companions close enough to suspect they could be responsible for the observed hole, 
namely sources \# 12, 27, 24, and 23, in increasing order of projected separation. 
Source \# 12 is a binary with a projected separation of $\sim$0.6 AU and $\lambda_{turn-off }$= 8.0 $\mu$m,
which implies the disk must be a circumbinary one. However, given the long $\lambda_{turn-off}$ wavelength 
and tightness of the binary system,  it seems unlikely that the present size of the inner hole
is directly connected to the presence of the stellar companion.
Source \#  27 is triple system with 
projected separations of $\sim$0.3 AU and 41 AU and $\lambda_{turn-off}$ = 4.5 $\mu$m. 
The IR excess could originate at  a circumbinary disk around  the tight components of the system or at a 
circumstellar disk around the wide component. 
Target \#24 is also a triple system with projected separations of $\sim$7 and 105 AU and $\lambda_{turn-off}$ = 5.8 
$\mu$m.  The IR excess could also in principle originate at  a circumbinary disk around  the tight components 
of the system.  However, a circumprimary disk around source \# 24-A is an equally likely possibility. 
Target \# 23 is  a  $\sim$0.19$''$ (or  24 AU) separation  binary with $\lambda_{turn-off}$ = 8.0 $\mu$m. 
This object has a small 24 $\mu$m excess, consistent with Wien side of the emission from  a cold outer disk. 
Objects \# 12 and  23 have a low disk masses ($< $ 1.7 M$_{JUP}$), negligible accretion, and  
L$_{D}$/L$_{*}$ $<$  10$^{-3}$.  As discussed in the previous section, they are likely to be in the 
debris disk stage, regardless of the origin of the inner hole.  
After target \#23, the next tighter binary is source \#5, with a projected separation of 0.54$''$ (or 
68 AU). Since this  object has  $\lambda_{turn-off}$ = 5.8 $\mu$m, suggesting the presence 
of dust at  $\lesssim$1 AU distances, it is very unlikely that the presence of its inner hole is 
related to the observed companion.  
Since VLT-AO observations are clearly not sensitive to companions with separations $\lesssim$ 8 AU, 
a complementary radial velocity survey to find the tightest companions would be highly desirable to 
identify additional circumbinary disks within our sample.  

\subsubsection{Grain growth}

Once primordial sub-micron dust grains grow into somewhat larger bodies (r $_{dust}$ $\gg$ $\lambda_{stellar-photons}$), most of the 
solid mass ceases to interact with the stellar radiation, and the opacity function decreases dramatically. 
Grain growth is a strong function of radius; it is more efficient in the inner regions where the surface density is higher 
and the dynamical timescales are shorter,  and hence can also produce opacity holes. 
Idealized dust coagulation models (i.e., ignoring fragmentation and radial drift) predict
extremely efficient grain growth (Dullemond $\&$ Dominik, 2005) resulting in the depletion
of all small grains in timescales of the order of 10$^{5}$ yrs, which is clearly inconsistent 
with the observational constraints on the lifetime of circusmtellar dust.  
In reality, however, the persistence of small opacity-bearing grains depends on a complex balance 
between dust coagulation and fragmentation (Dominik $\&$ Dullemond, 2008). 

\begin{figure}[t]
\includegraphics[width=6.0in]{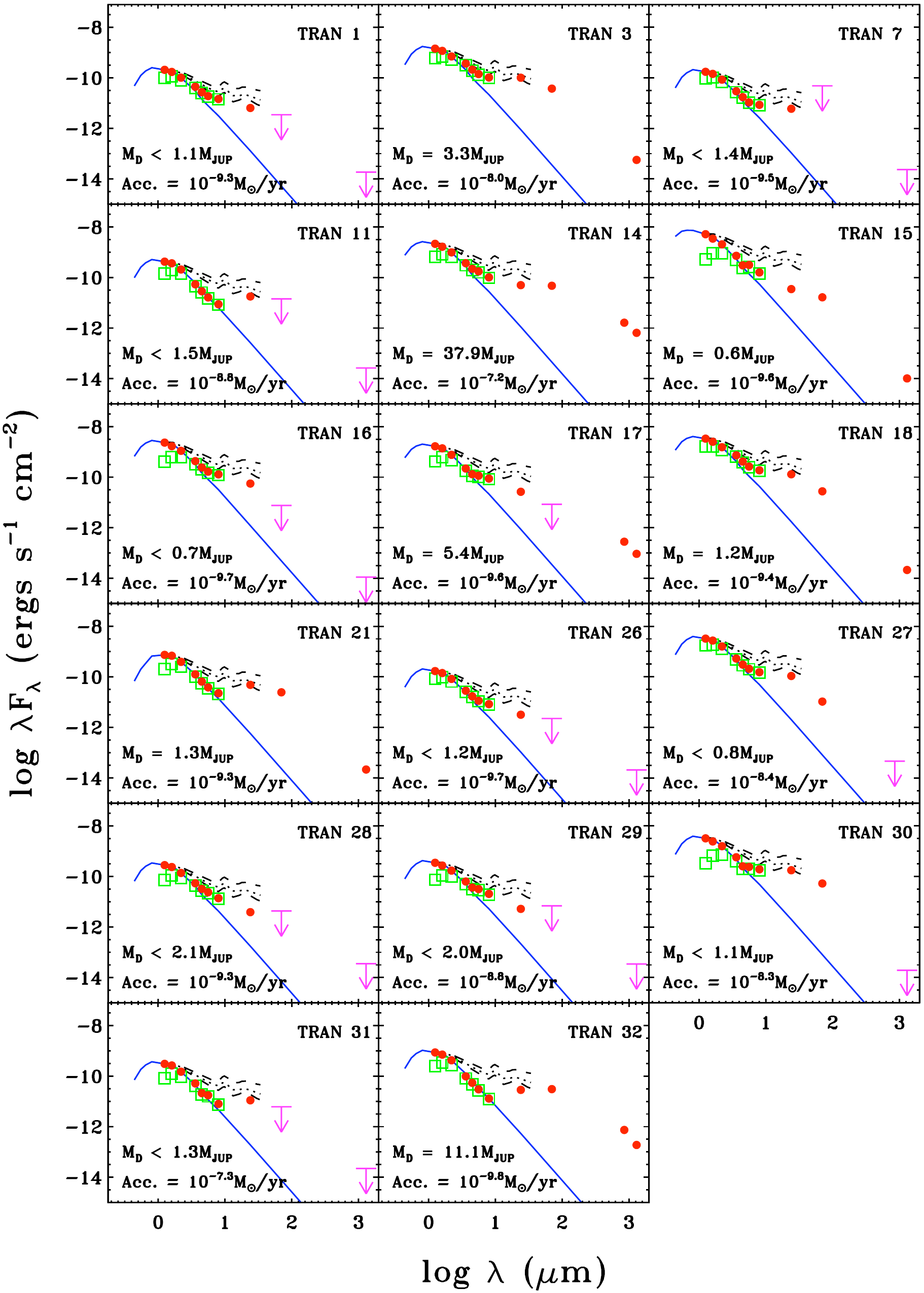}
\caption{The SEDs of all our accreting transition disk targets. 
The symbols are the same as in Fig. 9. Disk masses range from $<$ 0.7 to 
$\sim$40 M$_{JUP}$.  For targets \# 14, 17, and 32,  850 $\mu$m fluxes, from Andrews $\&$ Williams (2007) 
and Nutter et al. (1996),  are shown in addition to the 1.3 mm fluxes listed in Table 2. 
}
\label{sed_acc}
\end{figure}

Since grain growth affects only the dust and operates preferentially at smaller radii, a disk evolution dominated by 
grain growth is expected to result in an actively accreting  disk with reduced opacity in the inner regions.
All the accreting disks in our sample, whose SEDs are shown  in Figure~\ref{sed_acc}, are thus
in principle consistent with grain growth. 
Objects with very strong accretion (i.e.,  $\gtrsim$ 10$^{-8}$ M$_{\odot}$yr$^{-1}$) are especially good candidates for grain growth dominated 
disks as an opacity hole is likely to trigger the onset of the magneto-rotational instability and exacerbate accretion (Chiang $\&$ Murray-Clay, 2007). 
Target \#14, which is the extensively studied system DoAr 25, is a prime example of this scenario. 
It has, by far, the highest disk mass in our sample ($\sim$38 M$_{JUP}$) and one of the  highest accretion rates ($\sim$10$^{-7.2}$ 
M$_{\odot}$yr$^{-1}$).  This object has recently been imaged at high spatial resolution with the Submillimeter 
Array, and its SED and  visibilities  have been successfully reproduced with a simple model  
incorporating significant grain growth in its inner regions (Andrews et al. 2008). 

Grain growth is considered the first step toward planet formation. 
Unfortunately, the observational effects of grains growing into terrestrial
planets are no different from those of growing into meter size 
objects. As a result, the observations presented herein  place no
constraints on how far along the planet formation process currently is,
unless the planet becomes a giant planet, massive enough to
dynamically open a gap in the disk. 

\subsubsection{Giant planet formation} 

Since theoretical models of the dynamical interactions of forming giant
planets with the disk (Lin $\&$ Papaloizous 1979, Artymowicz $\&$ Lubow 1994) predict the formation of inner
holes and gaps, planet  formation quickly became one of the most exciting explanations proposed for the inner
holes of transition disks (Calvet et al. 2002, 2005; Quillen et al. 2004; D'Alessio et al. 2005; Brown et. 2008).
A planet massive enough to open a gap in the disk (M $\gtrsim$0.1-0.5  M$_{JUP}$), is expected to divert most of
the material accreting  from the outer disk onto itself.  As a result,  in the presence of  a Jupiter mass  planet, 
the accretion onto the star is expected to be reduced by a factor of 4 to 10 with respect to the accretion across the 
outer disk (Lubow $\&$ D$'$Angelo, 2006).

Four of the accreting objects in our sample, targets \# 11, 21, 31, and  32 have  $\alpha_{excess}$  $>$ 0,
suggesting the presence of a sharp inner hole, and are thus excellent candidates for ongoing \emph{giant} planet formation. 
Source \#32 is a particularly good candidate to be currently forming a massive giant planet, as it has 
a high disk mass ($\sim$11 M$_{JUP}$)  and small accretion rate (10$^{-9.8}$ M$_{\odot}$yr$^{-1}$). 
Sources \#11, 21, and 31  (M$_D$ $<$ 1.5 M$_{JUP}; $ M$_{acc}$ $\sim$10$^{-9.3}$-10$^{-7.3}$ M$_{\odot}$yr$^{-1}$), could have 
formed a massive giant planet in the recent past, as in their case most of the disk mass has already been depleted.
Since target \#21 has  been observed with the lunar occultation technique, an equal-mass 
binary  system can be ruled out for this system down to $\sim$1 AU (Simon et al. 1995). 
Also, it has been proposed that companions with masses of the order of 10 M$_{JUP}$   completely isolate the inner disk from the 
outer disk and halt the accretion onto the star (Lubow et al. 1999). If that is the case, accretion itself could be evidence
 \emph{against} the presence of a  close (sub)stellar companion.  That said, 
 because a few accreting close binary systems are already known, such as DQ Tau (Carr et al. 2001) and  CS Cha (Espaillat et al. 2007), 
 it is clear that accretion does not  rule out completely  the presence of tight stellar companions. 
 In addition to the mass of the companion,   the eccentricity of the orbit,  the viscosity and scale height of the disk are 
 also important factors in determining  whether or not gap-crossing streams can exist and allow the accretion onto the star 
 to continue (Artymowicz $\&$Lubow 1996). 
 
 \subsubsection{Disk Classification}
 
Based on the discussion above, we  divide our transition disk sample into
the following ``disk types": 
 
\noindent  \emph{a})   13  grain growth-dominated disks  (accreting objects with $\alpha$ $\lesssim$ 0). \\
\noindent  \emph{b})   4 giant planet forming disks (accreting objects with $\alpha$ $\gtrsim$ 0). \\
\noindent  \emph{c})   5  photoevaporating disks (non-accreting objects with disk mass  $<$ 2.5 M$_{JUP}$, 
but  L$_{D}$/L$_{*}$ $>$  10$^{-3}$)  \\
\noindent  \emph{d})   4 debris disks (non-accreting objects with disk mass  $<$ 2.5 M$_{JUP}$ and  L$_{D}$/L$_{*}$ $<$  10$^{-3}$) \\
\noindent  \emph{e})   4 circumbinary disks (a binary tight enough  to accommodate both components within the inferred inner hole). 

The total number of objects listed add up to 30 instead of 26 because 4 objects fall into two categories. 
Sources \# 12 and 23 are considered both debris disks and circumbinary disks. 
Object \#24 has been classified as both a circumbinary disk  and a photoevaporating  disk, while
object \# 27 has been classified  as both a  circumbinary  disk and a grain growth dominated disk. 
These last two objects are triple systems and their classification depends on whether the IR excess
is associated with  the tight components of the systems or with the wide components.

The ``disk types" for our targets are listed in Table 3. 
 All of our objects should be considered to be \emph{candidates} for the categories listed above. 
 The current classification represents our best  guess given the available data. 
 Only followup observations and detailed modeling will firmly establish the true nature
 of each object.  An important  caveat of this classification is the lack 
 of constraints for most of the sample on stellar companions within $\lesssim$ 8 AU, a range where  
 $\sim$30$\%$ of all stellar companions are expected to lay (Duquennoy $\&$ Mayor, 1991; Ratzka et  al. 2005). 
 Future radial velocity observations are very likely to increase the number of objects in the
 circumbinary disk category.

While it is tempting to speculate that  the  ``disk types"  \emph{a} through \emph{d} represent an evolutionary 
sequence,  starting  with a ``full disk" (i.e.,  a  ``full disk" that undergoes grain growth followed by giant 
planet formation, followed by  photoevaporation leading to a debris or disk-less stage)  it is clear that not 
all disks follow  this sequence.
For instance, many of the  grain growth dominated disks have disk masses that are 10 times smaller 
than some of the giant planet forming disks (e.g., targets \#1, 15, 16, and 26 vs. \#32).
These former systems have very little mass left in the disk ($\lesssim$ 1 M$_{JUP}$)  and 
 might never form giant planets.  It it also clear that not all disks become transition disks 
in the same way.  Some objects become transition disks (as defined by their SEDs)
while the total  mass  of the disk is very large (e.g.,  targets \#32 and 14, M$_D$ 
$\sim$10-40 M$_{JUP}$), but  other objects retain  ``full disks" even when the disk mass 
is relatively small (M$_D$ $\sim$2  M$_{JUP}$), such as ROXR1 29 (Cieza et al. 2008).  
Since  very little dust is needed to keep a disk optically thick at near and mid-IR wavelengths, it  
is  expected that some objects will only become transition disks when their disk masses and accretion 
rates are low enough to become susceptible to photoevaporation (i.e., they will evolve directly from a 
``full disk" to a photoevaporating disk followed by a debris or disk-less stage).

\section{Summary and Conclusions}

We have obtained millimeter wavelength photometry,  high-resolution
optical spectroscopy, and Adaptive Optics near-infrared imaging for a sample 
of  34 \emph{Spitzer}-selected  YSO candidates located in the Ophiuchus 
molecular cloud.  All our targets have SEDs consistent with circumstellar
disks with inner opacity holes (i.e., transition disks). 
After  removing one likely classical Be star and seven likely AGB stars we were left
with a sample of 26 transition disks.
We have used these  data to estimate  the disk mass, accretion rate, and multiplicity of each 
transition disk in our sample in order to  investigate the mechanisms potentially responsible for their 
inner opacity holes:  dynamical interaction with a stellar companion,  photoevaporation, grain 
growth,  and  giant planet formation. 

We find that transition disks  exhibit a wide range of masses, accretion rates, and SED morphologies. 
They clearly represent a heterogeneous group of objects, but overall,  transition disks
tend to have much lower masses and accretion rates than ``full disks." 
Eight of our targets are multiples: 6 are binaries and the other 2 are triple systems. 
In four cases, the stellar companions are close enough to suspect they  are 
responsible for the inferred inner holes. 
We do not  see an increased incidence of binaries with separations in the $\sim$8-20 AU range,
the  range where we are sensitive to  companions that could carve the inferred inner 
holes, suggesting companions at these separations are not responsible for a 
large fraction of the transition disk population.  
However,  given the small size of the current sample this result should not be overinterpreted. 
A complementary radial velocity survey to find the tightest companions is highly desirable to 
firmly establish the fraction of  transition disks that could be accounted for by very tight binaries.

We find that 9 of our transition disk targets  have low disk mass ($<$ 2.5 M$_{JUP}$) and negligible accretion 
($<$ 10$^{-11}$ M$_{\odot}$yr$^{-1}$), and are thus consistent with photoevaporating (or photoevaporated) disks. 
Four of the non-accreting objects have fractional disk luminosities $<$ 10$^{-3}$ and could already be in
the debris disk stage.
The remaining 17 objects are accreting.  Four of these accreting objects have SEDs 
suggesting the presence of sharp inner holes ($\alpha_{excess}$ values $\gtrsim$ 0), 
and thus are  excellent candidates for harboring giant planets.  
The other 13 accreting objects have $\alpha_{excess}$ values $\lesssim$ 0, 
which suggest a more or less radially continuous disk. These systems could be forming terrestrial planets, 
but their planet formation stage remains  unconstrained  by current observations.  

Understanding transition disks is key to understanding disk evolution and planet formation.  They are systems where 
important disk evolution processes such as grain growth, photevaporation, dynamical interactions and planet formation 
itself are clearly discernable. 
In the near future,  detailed studies of transition disks  such as sources \# 11, 21, 31, and 32,  will very likely revolutionize our 
understanding of planet formation.  In particular, the Atacama Large Millimeter Array (ALMA)  will have the resolution and 
sensitivity needed to image these transition disks, using both the continuum and  molecular tracers, at  $\sim$1-3 AU resolution.
Such exquisite observations will provide unprecedented observational constraints,  much needed to 
distinguish among competing theories of planet formation.
Finding promising targets for ALMA is one of the main goals of this paper, the first one of a series covering
over 100 \emph{Spitzer}-selected transition disks.

\acknowledgments
{\it{Acknowledgments:}} 
We thank the anonymous referee for very helpful comments that improved this paper. 
Support for this work was provided by NASA through
the \emph{Spitzer} Fellowship Program under an award from Caltech.
M.R.S thanks for support from FONDECYT (1061199) and Basal CATA PFB 06/09.
G.A.R. was supported by ALMA FUND Grant 31070021. M.D.M. was supported by ALMA-Conicyt
FUND Grant 31060010.
J.P.W. acknowledges support from the National Science Fundation Grant AST08-08144.
P.M.H. and N.J.E. thank the support from the \emph{Spitzer} Space Telescope 
Legacy Science Program, which was provided by NASA  through contracts 
1224608, 1230782, and 1230779, issued by the JPL/Caltech, under NASA contract 1407. 
This work makes use of  data obtained with the \emph{Spitzer} Space Telescope,
which is operated by JPL/Caltech,  under a contract with NASA. 

\emph{Facilities}:  {\it{Spitzer}} (IRAC, MIPS), SMA, CSO (Bolocam),  VLT (Conica), CLAY (Mike), 
CFHT (Espadons), Du Pont (Echelle).

\begin{deluxetable}{rrcrrrrrrrrrrr}
\rotate
\tablewidth{0pt}
\tabletypesize{\tiny}
\tablecaption{Transition Disk Sample}
\label{sample}
\tablehead{\colhead{\#}&\colhead{\emph{Spitzer} ID}&\colhead{Alter. Name}&\colhead{R$_1$}&\colhead{R$_2$}&\colhead{J\tablenotemark{a}}&\colhead{H}&\colhead{K$_{S}$}&\colhead{F$_{3.6}$\tablenotemark{a}}&\colhead{F$_{4.5}$}&\colhead{F$_{5.8}$}&\colhead{F$_{8.0}$}&\colhead{F$_{24}$}&\colhead{F$_{70}$\tablenotemark{b}  }   \\
\colhead{}&\colhead{}&\colhead{}&\colhead{(mag)}&\colhead{(mag)}&\colhead{(mJy)}&\colhead{(mJy)}&\colhead{(mJy)}&\colhead{(mJy)}&\colhead{(mJy)}&\colhead{(mJy)}&\colhead{(mJy)}&\colhead{(mJy)}& \colhead{(mJy)}   }
\startdata
1   & SSTc2d\_J162118.5-225458 & \nodata           & 15.50 & 15.81 & 4.21e+01 & 6.04e+01 & 5.81e+01 & 4.75e+01 & 3.75e+01 & 3.37e+01 & 3.77e+01 & 5.14e+01  &  $<$ 8.13e+01\\
2   & SSTc2d\_J162119.2-234229 &  HIP 80126    &    6.99 &   6.98 & 3.77e+03 & 2.56e+03 & 1.81e+03 & 7.81e+02 & 4.90e+02 & 3.74e+02 & 2.54e+02 & 1.92e+02  &   $<$1.02e+02 \\
3   & SSTc2d\_J162218.5-232148 &  V935 Sco      &  11.26 & 13.03 & 2.49e+02 & 3.76e+02 & 3.80e+02 & 3.83e+02 & 2.89e+02 & 2.47e+02 & 2.66e+02 & 8.08e+02  &          8.75e+02 \\
4   & SSTc2d\_J162224.4-245019 & \nodata           &  16.29 & 16.38 & 2.04e+02 & 4.87e+02 & 5.40e+02 & 3.05e+02 & 2.01e+02 & 1.73e+02 & 1.18e+02 & 3.09e+01  &  $<$ 2.45e+02 \\
5   & SSTc2d\_J162245.4-243124 & \nodata           &  14.34 & 14.82 & 1.13e+02 & 1.72e+02 & 1.58e+02 & 9.21e+01 & 6.15e+01 & 4.47e+01 & 5.13e+01 & 3.45e+02  &  $<$ 1.38e+02 \\
6   & SSTc2d\_J162312.5-243641 & \nodata           &  14.35 & 15.39 & 1.21e+02 & 2.49e+02 & 2.49e+02 & 1.33e+02 & 8.17e+01 & 6.49e+01 & 4.67e+01 & 1.62e+01  &  $<$ 1.27e+02 \\
7   & SSTc2d\_J162332.8-225847 & \nodata           &  16.37 & 16.22 & 4.03e+01 & 5.47e+01 & 5.09e+01 & 3.23e+01 & 2.44e+01 & 1.92e+01 & 2.23e+01 & 4.79e+01  &  $<$ 1.13e+03 \\
8   & SSTc2d\_J162334.6-230847 & \nodata           &  14.82 & 14.32 & 5.23e+02 & 1.11e+03 & 1.14e+03 & 6.30e+02 & 3.39e+02 & 3.18e+02 & 2.19e+02 & 7.73e+01  &  $<$ 1.92e+02 \\
9   & SSTc2d\_J162336.1-240221 &  \nodata          &  16.27 & 16.35 & 3.89e+01 & 5.94e+01 & 6.10e+01 & 5.92e+01 & 4.47e+01 & 3.94e+01 & 3.92e+01 & 4.78e+01  &  $<$ 2.51e+02 \\
10 & SSTc2d\_J162355.5-234211 & \nodata          &   16.26 & 16.37 & 5.72e+02 & 1.50e+03 & 1.90e+03 & 1.19e+03 & 7.48e+02 & 6.85e+02 & 4.59e+02 & 1.42e+02  &  $<$ 3.41e+02 \\
11 & SSTc2d\_J162506.9-235050 & \nodata          &   15.45 & 15.55 & 6.05e+01 & 1.05e+02 & 1.05e+02 & 5.44e+01 & 3.82e+01 & 2.81e+01 & 2.23e+01 & 1.42e+02  &  $<$ 3.35e+02 \\
12 & SSTc2d\_J162603.0-242336 & DoAr 21         &  11.07 &   9.34 & 9.26e+02  & 1.84e+03 & 2.15e+03 & 1.26e+03 & 8.78e+02  & 7.43e+02 & 6.89e+02 & 1.81e+03  &          1.20e+04 \\
13 & SSTc2d\_J162619.5-243727 & ROXR1 20    &   15.74& 15.80 & 5.32e+01 & 7.06e+01 & 6.06e+01 & 3.76e+01 & 2.88e+01 & 2.33e+01 & 2.63e+01 & 2.49e+01  &  $<$ 8.34e+02 \\
14 & SSTc2d\_J162623.7-244314 & DoAr 25         &   13.43 & 12.99 & 2.79e+02 & 4.48e+02 & 4.84e+02 & 3.67e+02 & 2.92e+02 & 2.99e+02 & 2.58e+02 & 3.99e+02  &          1.10e+03 \\
15 & SSTc2d\_J162646.4-241160 & ROXs 16       &    14.83 & 14.54 & 2.14e+02 & 4.87e+02 & 6.76e+02 & 6.03e+02 & 3.55e+02 & 4.95e+02 & 3.86e+02 & 2.78e+02  &          3.84e+02 \\
16 & SSTc2d\_J162738.3-235732 & DoAr  32       &    14.33 & 13.88 & 1.74e+02 & 3.34e+02 & 4.45e+02 & 3.95e+02 & 3.01e+02 & 2.72e+02 & 3.22e+02 & 4.46e+02  &  $<$ 1.76e+02  \\
17 & SSTc2d\_J162739.0-235818 & DoAr   33      &    13.78 & 13.88 & 1.75e+02 & 3.32e+02 & 3.48e+02 & 2.12e+02 & 1.69e+02 & 1.93e+02 & 2.22e+02 & 2.11e+02  &  $<$ 1.96e+02  \\
18 & SSTc2d\_J162740.3-242204 & DoAr    34     &    11.91 & 11.87 & 6.71e+02 & 8.78e+02 & 8.73e+02 & 7.84e+02 & 5.75e+02 & 4.65e+02 & 4.84e+02 & 1.04e+03  &          6.42e+02  \\
19 & SSTc2d\_J162802.6-235504 & \nodata         &     17.08 & 16.95 & 3.16e+01 & 5.56e+01 & 5.77e+01 & 2.61e+01 & 1.41e+01 & 1.47e+01 & 1.30e+01 & 4.84e+01  &  $<$ 1.41e+02  \\
20 & SSTc2d\_J162821.5-242155 & \nodata         &     17.49 & 17.34 & 2.35e+01 & 5.17e+01 & 5.64e+01 & 3.61e+01 & 2.51e+01 & 1.75e+01 & 1.06e+01 & 3.78e+00  &  $<$ 1.18e+02  \\
21 & SSTc2d\_J162854.1-244744 & WSB 63        &     15.74 & 15.41 & 8.50e+01 & 1.73e+02 & 1.83e+02 & 1.23e+02 & 8.46e+01 & 6.50e+01 & 5.69e+01 & 3.84e+02  &          5.68e+02  \\
22 & SSTc2d\_J162923.4-241357 & \nodata         &     17.14 & 16.39 & 1.96e+02 & 5.77e+02 & 1.04e+03 & 1.21e+03 & 9.22e+02 & 8.11e+02 & 5.14e+02 & 2.06e+02  &  $<$ 8.44e+01  \\
23 & SSTc2d\_J162935.1-243610 & \nodata         &     16.94 & 17.40 & 4.95e+01 & 8.62e+01 & 8.92e+01 & 5.54e+01 & 4.22e+01 & 3.18e+01 & 2.33e+01 & 6.64e+00  &  $<$ 7.39e+01  \\
24 & SSTc2d\_J162944.3-244122 & \nodata         &     14.25 & 14.71 & 1.48e+02 & 1.65e+02 & 1.47e+02 & 9.05e+01 & 6.33e+01 & 4.42e+01 & 3.46e+01 & 4.20e+01  &  $<$ 1.13e+02  \\
25 & SSTc2d\_J163020.0-233108 & \nodata         &     15.62  & 15.47 & 4.72e+01 & 7.33e+01 & 6.54e+01 & 3.81e+01 & 2.72e+01 & 2.10e+01 & 1.79e+01 & 1.03e+01  &  $<$ 6.55e+01 \\
26 & SSTc2d\_J163033.9-242806 & \nodata         &     16.17 & 16.33 & 3.57e+01 & 5.08e+01 & 4.79e+01 & 3.02e+01 & 2.36e+01 & 2.01e+01 & 2.16e+01 & 2.53e+01  &  $<$ 5.23e+01 \\
27 & SSTc2d\_J163115.7-243402 & V2131 Oph  &     10.92 & 10.99 & 7.24e+02 & 1.01e+03 & 9.38e+02 & 5.75e+02 & 4.28e+02 & 3.71e+02 & 3.97e+02 & 8.62e+02  &          2.44e+02 \\
28 & SSTc2d\_J163145.4-244307 & \nodata         &      17.94 & 17.94 & 3.05e+01 & 5.79e+01 & 6.34e+01 & 5.22e+01 & 4.17e+01 & 4.02e+01 & 3.49e+01 & 3.10e+01  &  $<$ 1.00e+02 \\
29 & SSTc2d\_J163154.4-250349 & \nodata         &      17.12 & 17.17 & 3.09e+01 & 5.89e+01 & 7.32e+01 & 5.95e+01 & 4.64e+01 & 5.27e+01 & 5.15e+01 & 4.17e+01  &  $<$ 1.61e+02 \\
30 & SSTc2d\_J163154.7-250324 & WSB 74        &      14.88 & 15.08 & 1.40e+02 & 3.52e+02 & 5.30e+02 & 4.84e+02 & 2.98e+02 & 3.78e+02 & 4.68e+02 & 1.41e+03  &          1.23e+03 \\
31 & SSTc2d\_J163205.5-250236 & WSB  75       &      16.94 & 15.95 & 3.48e+01 & 6.67e+01 & 7.00e+01 & 5.03e+01 & 2.78e+01 & 2.97e+01 & 2.02e+01 & 8.84e+01  &  $<$ 1.43e+02 \\
32 & SSTc2d\_J163355.6-244205 & RXJ1633.9-2242  & 14.67 & 15.04 & 1.04e+02 & 1.85e+02 & 2.02e+02 & 9.67e+01 & 7.10e+01 & 5.12e+01 & 3.28e+01 & 2.28e+02  &          7.13e+02 \\
33 & SSTc2d\_J163603.9-242344 & \nodata                    & 16.74 & 16.51 & 7.89e+01 & 1.72e+02 & 1.87e+02 & 1.13e+02 & 7.46e+01 & 6.05e+01 & 4.33e+01 & 1.22e+01  &  $<$ 8.03e+01 \\
34 & SSTc2d\_J164429.3-241555 & \nodata                    & 18.23 & 17.57 & 2.96e+02 & 6.06e+02 & 7.52e+02 & 4.60e+02 & 3.08e+02 & 2.54e+02 & 1.77e+02 & 7.57e+01  &  $<$ 5.76e+01
\enddata
\tablenotetext{a}{All the 2MASS, IRAC and 24 $\mu$m  detections are $\ge$7-$\sigma$ (i.e.,the photometric uncertainties are $\lesssim$15$\%$)} 
\tablenotetext{b}{$\ge$5-$\sigma$ detections from the \emph{Cores to Disks} catalogs or  5-$\sigma$ upper limits as  described in \S~\ref{sed_mor}.} 
\end{deluxetable}

\begin{deluxetable}{rrrccccrrrrrrrr}
\rotate
\tablewidth{0pt}
\tabletypesize{\scriptsize}
\tablecaption{Observed Properties}
\label{observed}
\tablehead{\colhead{\#}&\colhead{Ra (J2000)}&\colhead{Dec (J2000)}&\colhead{Tel. SpT}&\colhead{SpT.}&\colhead{Li I\tablenotemark{a}}&\colhead{Ca II\tablenotemark{a}}&\colhead{H$\alpha$\tablenotemark{b}}&\colhead{$\lambda_{mm}$}&\colhead{Flux$_{mm}$\tablenotemark{c}}&\colhead{$\sigma$Flux$_{mm}$}&\colhead{Separ\tablenotemark{d}}&\colhead{pos. ang.}&\colhead{$\Delta$K} \\
\colhead{}&\colhead{(deg)}&\colhead{(deg)}&\colhead{}&\colhead{}&\colhead{(\AA)}&\colhead{}&\colhead{(km/s)}&\colhead{(mm)}&\colhead{(mJy)}&\colhead{(mJy)}&\colhead{(arcsec)}&\colhead{(deg)}&\colhead{(mag) }}
\startdata
1  & 245.32697  & -22.91608  &   Clay       &   M2   &      0.48      &    Yes     &  363         &  1.30   &  $<$  8.10  &  \nodata  &  \nodata  &  \nodata  & $>$  2.85 \\
2  & 245.32991  & -23.70796  &   CFHT     &   B5  &     \nodata   &    No     &  597           &  1.30   &  $<$ 4.80    &  \nodata  &  \nodata  &  \nodata  &  $>$ 3.17 \\
3  & 245.57717  & -23.36337  &   CFHT     &   K5  &       0.47      &    Yes    &  493          &  1.30    &        24.50   &  3.10        &  \nodata  &  \nodata  &  $>$ 3.23 \\
4  & 245.60170  & -24.83854  &   Du Pont &   M5 &    \nodata   &    No      &     -1          &  1.10   &$<$ 16.30    &  \nodata  &  \nodata  &  \nodata  &  $>$ 3.43 \\
5  & 245.68912  & -24.52328  &   CFHT     &  M3  &      0.38       &    Yes    &  150          &  1.30   & $<$  6.30    &  \nodata  &   0.54       &    35          &  0.12 \\
6 & 245.80225   & -24.61147  &   Du Pont &  M2  &    \nodata   &    No      &      -1         & 1.30    & $<$ 5.40     &  \nodata  &  \nodata  &  \nodata  &  $>$ 3.11 \\
7  & 245.88680  & -22.97967  &  Clay        &   M5 &      0.56       &    Yes     & 344           &  1.30   & $<$ 10.20   &  \nodata  &  \nodata   &  \nodata  &  $>$ 2.69 \\
8  & 245.89427  & -23.14627  &   CFHT     &   M5 &     No           &    No      &      -1          &  1.30   & $<$ 5.80     &  \nodata  &  \nodata   &  \nodata  &  $>$ 2.72 \\
9  & 245.90040  & -24.03915  &  Clay        &   M5 &      0.64        &   Yes      &  146           &  1.30   & $<$ 11.40  &  \nodata  &  1.68         & 144          &  0.60 \\
10& 245.98142  & -23.70292  &  Du Pont &   M5 &     \nodata    &   No       &     -1           & 1.10    & $<$ 9.60     &  \nodata  &  \nodata  &  \nodata  &  $>$ 3.18 \\
11& 246.27877  & -23.84730  &  Clay       &   M3  &      0.56        &   Yes      &  414           &  1.30  & $<$11.40     &  \nodata  &  \nodata  &  \nodata  &  $>$ 3.23 \\
12& 246.51255  & -24.39334  &   CFHT    &   K1 &      \nodata    &   Yes     &   No    &  0.85  & $<$ 18.00    &  \nodata  &    0.005      &  \nodata  & \nodata \\
13& 246.58117  & -24.62429  &  Clay       &   M5 &      0.58         &\nodata &  165           &  1.30  & $<$ 4.30      &  \nodata  &  \nodata  &  \nodata  &  $>$ 3.06 \\
14& 246.59863  & -24.72055  & CFHT     &    K5 &      0.48         &   Yes      &   573          &  1.30  &280.00          & 10.00      &  \nodata   &  \nodata  &  $>$ 2.96 \\
15& 246.69341  & -24.19997  &  CFHT    &    G5  &     0.56         &  Yes       &   330         &  1.30  &         4.5        &  1.6          &  0.55         &  249        &  1.76          \\
16& 246.90971  & -23.95893  &  CFHT    &    K5&       0.52         &  Yes       &   322          &  1.30  & $<$ 4.80     &  \nodata  &  \nodata  &  \nodata  &  $>$ 2.70 \\
 17& 246.91254  & -23.97174  & CFHT    &    K6 &      0.44         &  Yes       &    329         &  1.30  & 40.00           &  10.00      &  \nodata  &  \nodata  &  $>$ 3.20 \\
 18& 246.91778 & -24.36777  &CFHT      &    K5 &      0.50         &  Yes       &    351         &  1.30  &  9.20            &  2.71        &  0.65        &   4.97      &   2.70 \\
 19& 247.01074  & -23.91767& Clay        &    M3  &     0.62          &   Yes      &    159        &  1.30  & $<$ 9.90      &  \nodata  &  \nodata  &  \nodata  & $>$  3.33 \\
  20& 247.08958 & -24.36525& Clay        &    M3 &     0.57         &   Yes       &      160       &  1.10  & $<$ 16.50   &  \nodata  &  \nodata   &  \nodata  &  $>$ 3.19 \\
  21& 247.22524 & -24.79563 &Du Pont  &    M2 &     \nodata    &  Yes       &     365          &  1.30  &  9.30            &  3.00       &  \nodata    &  \nodata  &  $>$ 3.15 \\
  22  & 247.34741  & -24.23241& Clay      &    M6&      No      &   No       &     No     &  1.30  & $<$ 5.40     &  \nodata  &  \nodata  &  \nodata  &  $>$ 2.40 \\
  23  & 247.39620  & -24.60289&  Clay     &    M4   &    0.49          &  \nodata& 199             &  1.10  & $<$ 17.40  &  \nodata  &  0.19 &  242  &  0.24 \\
  24  & 247.43449  & -24.68938&  CFHT  &    M4 &      \nodata      & Yes        &   152          &  1.30  & $<$ 15.30  &  \nodata  &  0.84 &  100  &  0.04 \\
  25  & 247.58349  & -23.5189 & Du Pont &   M4  &     \nodata    &  Yes         &       229        & 1.30  & $<$ 4.80      &  \nodata  &  \nodata  &  \nodata  &  $>$ 2.86 \\
 26  & 247.64124  & -24.46840& Clay       &    M4    &   0.42            & \nodata  &   318          &  1.30  & $<$ 9.00  &  \nodata  &  \nodata  &  \nodata  & $>$  2.87 \\
 27  & 247.81556  & -24.56724 &CFHT    &    K6  &       0.36            &Yes          &    450        &  0.85  & $<$ 13.00    &  \nodata  &   0.33  &  203   & 1.60 \\
  28  & 247.93913  & -24.71867  &  Clay   &    M4    &     0.51          &  Yes         &   356        &  1.30  & $<$ 15.30  &  \nodata  &  \nodata  &  \nodata  &  $>$ 3.11 \\
 29  & 247.97673  & -25.06370  &  Clay   &      M4    &   0.32          &  Yes        & 414        &  1.30  & $<$ 15.00  &  \nodata  &  \nodata  &  \nodata  & $>$  3.30 \\
  30  & 247.97805  & -25.05666 &Du Pont&  K7     &\nodata         &   Yes       &  470        &  1.30  &  $<$ 8.40  &  \nodata  &  \nodata  &  \nodata  &  $>$ 2.60 \\
  31  & 248.02306  & -25.04340  & Clay  &   M2     &    0.51             &  Yes       &  567        &  1.30  & $<$ 9.60  &  \nodata  &  \nodata  &  \nodata  &  $>$ 3.10 \\
  32  & 248.48165  & -24.70138  &  Du Pont&  K7    &  0.48           & Yes        &   301        &  1.30  & 81.80  &  2.70    &  \nodata  &  \nodata  & $>$  3.01 \\
  33  & 249.01642  & -24.39565  &  Du Pont& M4& \nodata            & \nodata   &   -1           &   1.10  & $<$ 18.00  &  \nodata  &  \nodata  &  \nodata  &$>$   3.51 \\
  34  & 251.12216  & -24.26541  &   Clay      & M6&   No          &  \nodata  &    -1           &  1.30  & $<$ 6.00  &  \nodata  &  \nodata  &  \nodata  & $>$  3.39 \\
 \enddata
\tablenotetext{a}{``\nodata" implies that the signal to noise in this region of the spectrum is too low to measure the width or establish the presence of the line}.
\tablenotetext{b}{``-1" implies that H$\alpha$  is seen in absorption.}
\tablenotetext{c}{ The 1.3 mm data  for source \# 14 and 17 comes 
from Andrews $\&$ Williams (2007). The 1.3 mm and  850 $\mu$m data for source \# 12, 13 and 27 comes
from Cieza et al. (2008).}
\tablenotetext{d}{ Source \# 12 is a binary identified by VLBA observations (Loinard et al. 2008).
 Source \# 24 is a triple system. The tight components are consistent with two equal-brightness objects with  a separation of $\sim$0.05$''$ and a $\sim$30 deg position angle
(see \S~\ref{multi_sec} and Figure ~\ref{multi}).
Source \# 27  is a triple system.  The ``primary" star in the VLT observations is itself 
a spectroscopic binary with a 35.9 d period (Mathieu et al. 1994).} 
\end{deluxetable}

\begin{deluxetable}{lrrrrrrrc}
\label{derived}
\tabletypesize{\scriptsize}
\tablewidth{0pt}
\tablecaption{Derived Properties}
\tablehead{\colhead{\#}&\colhead{LOG(Acc. rate)}&\colhead{Mass Disk\tablenotemark{a}}&\colhead{r$_{proj.}$\tablenotemark{b}}&\colhead{$\lambda_{tun-off}$}&\colhead{$\alpha_{excess}$}&\colhead{LOG(L$_{D}$/L$_{*}$)}&
\colhead{$A_J$}&\colhead{Object  Type}\\
\colhead{}&\colhead{(M$_{\odot}$/yr)}&\colhead{(M$_{JUP}$)}&\colhead{(AU)}&\colhead{$\mu$m}&\colhead{}&\colhead{}&\colhead{(mag)}&\colhead{}
}
\startdata
 1  &      -9.3  &  $<$  1.1  &  \nodata  &  2.20  & -1.16  &  -1.88  &   0.7 & grain growth dominated  disk  \\ 
 2  &       \nodata  &    \nodata  &  \nodata  &  8.00  & -1.25  & $<$ -4.09  & 0.2 &   Be star \\
 3  &      -8.0  &   3.3  &  \nodata  &  2.20  & -0.81  & -1.68  &   1.1 &  grain growth dominated  disk \\ 
 4  &      \nodata  &  \nodata &  \nodata  &  8.00  & -2.28  & -3.92  &  1.7 &   AGB star \\ 
 5  &    $<$  -11.0  &  $<$  0.8  &  68  &  5.80  &  0.38  & -2.32  &  0.7 &   photoevaporating  disk\\
 6  &       \nodata  &  \nodata &  \nodata  &  8.00  & -2.01  & $<$ -3.92  &  1.3 &   AGB star  \\ 
 7  &      -9.5  &  $<$  1.4  &  \nodata  &  4.50  & -0.63  & -2.55  &   0.4 &  grain growth dominated disk  \\ 
 8  &     \nodata  &  \nodata &  \nodata  &  8.00  & -2.00  & $<$ -3.70  &  1.4 &   AGB star \\
 9  &     $<$ -11.0  &  $<$ 1.5  &   210  &  2.20  & -1.20  & -1.50  & 0.9 &   photoevaporating disk\\
10  &    \nodata  &  \nodata &  \nodata  &  8.00  & -2.14  & $<$ -3.72  &  2.0 & AGB star  \\
11  &      -8.8  &  $<$  1.5  &  \nodata  &  8.00  &  0.65  &  $<$ -2.72  &  1.0 &   giant planet forming disk \\
12  &    $<$  -11.0  &  $<$  1.1  &  0.62 &  8.00  & -0.21  & -3.96  &  2.0 &   circumbinary/debris disk \\
13  &     $<$  -11.0  &  $<$  0.6  &  \nodata  &  5.80  & -0.99  & -2.50  & 0.1 &   photoevaporating disk\\
14  &      -7.2  &  37.5  &  \nodata  &  2.20  & -1.25  & -2.06  &  1.2 &  grain growth dominated disk\\
15  &      -9.6  &  0.6  &  69  &  4.50  & -1.27  & -2.51  &   2.7 &  grain growth dominated disk \\
16  &      -9.7  &  $<$  0.7  &  \nodata  &  2.20  & -1.25  & -2.01  &  1.9 & grain growth dominated disk\\
17  &      -9.6  &   5.4  &  \nodata  &  4.50  & -0.95  & -2.44  & 1.4 &    grain growth dominated disk\\
18  &      -9.4  & 1.2  &  81  &  2.20  & -1.02  & -1.78  &  0.8 &   grain growth dominated disk \\
19  &    $<$  -11.0  &  $<$  1.3  &  \nodata  &  8.00  &  0.16  & $<$ -2.96  &  1.1 &   photoevaporating disk\\
20  &    $<$  -11.0  &  $<$  1.6  &  \nodata  &  8.00  & -1.99  & $<$ -3.21  &   1.6 &  debris disk \\
21  &      -9.3  &   1.3  &  \nodata  &  8.00  &  0.69  & -2.30  &  1.4 &  giant planet-forming disk \\
 22  &   \nodata  & \nodata &  \nodata  &  8.00  & -1.92  & $<$ -3.03  &  2.6 &   AGB star  \\
23  &     $<$ -11.0  &   $<$1.7  &  24   &  8.00  & -2.18  &  $<$ -3.76  &   1.0 &  circumbinary/debris disk\\
24  &     $<$ -11.0  &   $<$2.1  &  105  &  5.80  & -1.07  & -2.85  &  0.0 &   circumbinary/photoeva. disk\\
25  &     $<$ -11.0  &  $<$  0.7  &  \nodata  &  5.80  & -1.56  & -3.41  &  0.7 &   debris disk  \\
26  &      $<$ -9.7  &  $<$  1.2  &  \nodata  &  4.50  & -1.00  & -2.44  &    0.6  &  grain growth dominated disk\\
27  &      -8.4  &   $<$ 0.8  &  41 &  4.50  & -0.62  & -2.38  &   0.7 &            circum./grain growth dominated disk  \\
 28  &      -9.3  &  $<$  2.1  &  \nodata  &  2.20  & -1.49  & -2.06  &  1.3 &   grain growth dominated disk \\
29  &      -8.8  &  $<$  2.0  &  \nodata  &  3.60  & -1.31  & -2.24  &   1.6 &  grain growth dominated  disk\\
30  &      -8.3  &  $<$  1.1  &  \nodata  &  4.50  & -0.21  & -2.23  &    2.6 &  grain growth dominated  disk\\
 31  &      -7.3  &  $<$  1.3  &  \nodata  &  8.00  &  0.30  & $<$ -2.73  &   1.3  &  giant planet forming disk\\
32  &      -9.9  &          11.1  &  \nodata  &  8.00  &  0.72  & -2.70  & 1.3 &    giant planet  forming disk \\
33  &     \nodata  & \nodata  &  \nodata  &  8.00  & -2.21  & $<$ -3.86  &  1.6 &    AGB star    \\
34  &     \nodata  &  \nodata  &  \nodata  &  8.00  & -1.83  & $<$ -3.58  &   1.4 &   AGB star   
\enddata
\tablenotetext{a}{The disk mass upper limits for targets  \#12 and 27 come from Cieza et al. (2008).}
\tablenotetext{b}{ 
Source \# 12 is a binary identified by VLBA observations (Loinard et al. 2008).
Source \# 24 is a triple system. The tight components are consistent with two equal-brightness objects with  a separation of $\sim$7 AU and a $\sim$30 deg position angle
(see \S~\ref{multi_sec} and Figure ~\ref{multi}).
Source \# 27  is a triple system.  The ``primary" star in the VLT observations is itself 
a spectroscopic binary with a 35.9 d period and estimated separation of 0.27 AU (Mathieu et al. 1994).} 
\end{deluxetable}

\end{document}